\documentclass[lettersize,journal]{IEEEtran}
\usepackage{amsmath,amsfonts}
\usepackage{algorithmic}
\usepackage{algorithm}
\usepackage{array}
\usepackage[caption=false,font=footnotesize,labelfont=rm,textfont=rm]{subfig}
\usepackage{textcomp}
\usepackage{stfloats}
\usepackage{url}
\usepackage{verbatim}
\usepackage{graphicx}
\usepackage{cite}
\usepackage{cuted}
\begin{document}

\title{Optimal Reference Nodes Deployment for Positioning Seafloor Anchor Nodes}

\author{Wei Huang,~\IEEEmembership{Member,~IEEE,} Pengfei Wu, Tianhe Xu, Hao Zhang,~\IEEEmembership{Senior Member,~IEEE,} Kaitao Meng*,~\IEEEmembership{Member,~IEEE,}
\thanks{Wei Huang, Pengfei Wu and Hao Zhang are with the Faculty of Information Science and Engineering, Ocean University of China, Qingdao, China (emails: hw@ouc.edu.cn, wupengfei@stu.ouc.edu.cn, zhanghao@ouc.edu.cn).}
\thanks{Tianhe Xu is with the School of Space Science and Physics, Shandong University (Weihai), Weihai, China (email:thxu@sdu.edu.cn).}
\thanks{ Kaitao Meng is with the Department of Electronic and Electrical Engineering, University College London, London, UK (email: kaitao.meng@ucl.ac.uk).}
\thanks{Corresponding author: Kaitao Meng.}
\thanks{Manuscript received XX XX, 2024; revised XX XX, 2024.}}

\markboth{IEEE Transactions on Communications,~Vol.~XX, No.~XX, August~2024}%
{Wei Huang \MakeLowercase{\textit{et al.}}: Optimal Reference Nodes Deployment for Positioning Underwater Anchors}


\maketitle

\begin{abstract}
Seafloor anchor nodes, which form a geodetic network, are designed to provide surface and underwater users with positioning, navigation and timing (PNT) services. Due to the non-uniform distribution of underwater sound speed, accurate positioning of underwater anchor nodes is a challenge work. Traditional anchor node positioning typically uses cross or circular shapes, however, how to optimize the deployment of reference nodes for positioning underwater anchor nodes considering the variability of sound speed has not yet been studied. This paper focuses on the optimal reference nodes deployment strategies for time--of--arrival (TOA) localization in the three-dimensional (3D) underwater space. We adopt the criterion that minimizing the trace of the inverse Fisher information matrix (FIM) to determine optimal reference nodes deployment with Gaussian measurement noise, which is positive related to the signal propagation path. A comprehensive analysis of optimal reference-target geometries is provided in the general circumstance with no restriction on the number of reference nodes, elevation angle and reference-target range. A new semi-closed form solution is found to detemine the optimal geometries. To demonstrate the findings in this paper, we conducted both simulations and sea trials on underwater anchor node positioning. Both the simulation and experiment results are consistent with theoretical analysis.
\end{abstract}


\section{Introduction}
\IEEEPARstart{A}coustic positioning is a fundamental technology in ocean observation, and high--precision positioning is crucial because marine environmental monitoring, disaster warning, and autonomous underwater vehicle (AUV) navigation require accurate positioning information. A seafloor geodetic network is a kind of underwater system that provides positioning, navigation and timing (PNT) services to surface and underwater users\cite{Yuanxi2017ProgressesAP,Yang2020Seafloor}, or can be deployed in many marine disaster warning and monitoring applications, including tsunami, seafloor spreading and submarine earthquakes\cite{Chadwell1996Precision,Chadwick2002Observatory,Blum2010Seafloor}. The seafloor geodetic network typically consists of a series of geodetic stations, also known as seafloor anchor nodes, deployed on the seafloor to form an underwater positioning system. This network works in a similar way to the Global Navigation Satellite System (GNSS) constellation. In the above applications, accurate underwater acoustic positioning information of objects is critical, so it is important to ensure the positioning accuracy of these geodetic stations.

\indent GNSS Acoustic (GNSS--A) technology has been widely adopted to locate the seafloor base nodes due to its precise horizontal positioning capability with a localization accuracy of even less than 1 decimeter \cite{SPIESS1998GAcoustic}. For GNSS--A technology, the main research areas include reliable data transmission and distance estimation \cite{Rodionov2020Modem,Khazan2023Modem}, accurate and real--time observation of the ocean sound speed field (SSF) distribution \cite{Huang2021SSPInversion,Huang2023Meta,Wang2024LSSVM}, and optimized deployment of reference nodes \cite{David2013OptimalPlacement,Han2015Deployment,Nguyen2016TOA,David2016OptimalPlacement,Xu2017OptimalAOA,Xu2019OptimalTOA,Zhao2012OptimalSP,Rui2014Elliptic}. 

\indent The high precision measurement of signal propagation time is mainly determined by the performance of the transducer, such as the stability of the equipment, the design of the signal, and so on. Various stable transducers based on digital signal processors (DSPs) or field-programmable gate arrays (FPGAs) have been produced over the last few decades \cite{Kulik2018Modems,Rodionov2020Modems}. The instrumental accuracy of underwater acoustic equipment at distances of 10 km can reach units of centimeters in a stable underwater acoustic channel with weak multipath effects. 

\indent The accurate and real--time observation of ocean SSF distribution can not only be measured through instruments such as sound speed profiler (SVP) or conductivity, temperature, depth profiler (CTD), but can also be quickly and accurately estimated through intelligent SSF construction methods \cite{Huang2021SSPInversion,Huang2023Meta}. Based on the accurately measured or estimated sound speed profile (SSP), the real signal propagation path under the influence of Snell effect can be reconstructed via ray tracing technique, so as to improve the underwater positioning accuracy, such as \cite{Liu2016JSL,Zhang2017Stratification,Huang2024Localization}. In \cite{Liu2016JSL,Zhang2017Stratification}, two joint localization and time synchronization algorithms were proposed for accurate positioning, which all combined the ray tracing technique. In \cite{Huang2024Localization}, an iterative localization algorithm was proposed considering the ray tracing problem when the target depth is not known. These works assumed that the reference nodes were uniformly deployed and the impact of reference node deployment on the accuracy of localization has not been discussed. However, the performance of network localization is greatly affected by the deployment of reference nodes, so how to optimally deploy nodes for accurately positioning seafloor anchor nodes needs to be tackled, and it is an open challenging issue.

\indent In addition to the industrial and technological developments, optimizing the deployment and selection of reference nodes has become a hot research topic to improve the accuracy performance of positioning systems \cite{Fawcett1988RangeEstimation,Oshman1999Localization}, as it affects the geometric dilution precision (GDOP) similar to the GNSS. Different optimization methods have been proposed to evaluate the estimation performance \cite{Adrian2010Optimality,Sonia2006Optimal,Pan2019Novel}. The Cram\'er-Rao lower bound (CRLB) was first proposed in \cite{Fawcett1988RangeEstimation} that has become a commonly evaluation standard applied for target estimation performance. Then how the CRLB and Fisher information matrix (FIM) affect the estimation performance is described in \cite{Oshman1999Localization}. Based on the analysis of FIM and CRLB, previous works usually focus on optimal sensor placement for two-dimension (2D) problem. While in the three-dimension (3D) ocean space, it becomes more challenging that the FIM increases to the scale of 3 × 3, leading to a more cumbersome problem. In \cite{Zhao2012OptimalSP}, a D-optimality criterion, that maximizing the determinant of FIM, was applied and a unified optimal sensor placement solution was derived in both the 2D and 3D circumstances. In \cite{Rui2014Elliptic}, the author proposed an optimal geometry analysis for 2D and 3D time--of--arrival (TOA) localization based on an A-optimality criterion (minimizing the trace of inverse FIM). However, the prerequisite is that the nodes are evenly distributed and the distribution of measurement noise is the same.
Based on the range-only measurement, in \cite{David2016OptimalPlacement}, the D-optimality criterion was applied to derive a general characterization of the optimal sensor configurations under the uniform noise variance assumption for the 3D underwater target localization. But it failed to obtain a closed-form solution. To tackle this issue, Xu et al. \cite{Xu2019OptimalTOA} proposed an optimal reference node deployment scheme for target localization in 3D scenes under time of arrival (TOA) measurement. In recent ocean research work, surface reference nodes are usually deployed in circular or cross shapes \cite{WANG2021KF,Xue2022GNSSA}. Both \cite{WANG2021KF} and \cite{Xue2022GNSSA} considered the stratification compensation (known as sound speed error correction) and compared the positioning performance between circle and cross shapes. The result indicates that in the circle case constraint information or regularization on the height coordinate is necessary to stabilize the differential solution. 

\indent Different from open 3D space, the scene for positioning seafloor anchor node becomes more unique because the reference nodes are usually located on the sea surface and the target nodes are located on the seafloor. The work of \cite{Zhao2012OptimalSP,Rui2014Elliptic,David2016OptimalPlacement,Xu2019OptimalTOA} can not be directly applied for localizing the seafloor anchor node because of the following three reasons. Firstly and most importantly, the uneven distribution of underwater sound speed causes the signal propagation path to bend due to the Snell effect, hence the TOA measurement can not reflect the actual light of sight distance. The phenomenon of signal refraction makes underwater positioning fundamentally different from terrestrial positioning. Secondly, the measurement error of the TOA is usually positively correlated with the actual propagation distance, and the bending of the sound lines makes it difficult to evaluate the measurement error of the light of sight TOA. Thirdly, the topological constraint relationship has changed compared to traditional 3D scenes. This means that some conclusions are no longer applicable. While for node deployment work of \cite{WANG2021KF,Xue2022GNSSA}, although the stratification compensation is considered, they ignored the fact that the distance measurement error is proportional to the real signal propagation distance, and there is still a lack of theoretical basis for which deployment strategy is optimal and how to set parameters such as radius.  

\indent In this paper, we focus on the optimal reference node deployment scheme for positioning seafloor anchor node, in response to the impact of ocean 3D positioning environment and non--uniform distribution of sound speed. In addition, the error analysis is performed from a more practical perspective, where the magnitude of the measurement error is proportional to the true signal propagation path. The contributions of this article can be concluded as follows:

\begin{enumerate}
	\item We considered a more practical distribution of the underwater range error, where the error is proportional to the actual signal propagation path. Based on the ray tracing theory, we found that the equivalent light--of--sight error still follows a Gaussian distribution.
	\item We analyzed the seafloor anchor node positioning scene considering the Snell effect, and optimized the deployment of reference nodes by optimizing the Fisher information matrix without constraints on the number of reference nodes and the distance from the target to reference nodes. Meanwhile, the Cram\'er-Rao Lower Bound (CRLB) is derived.
	\item To simplify the problem, we considered a situation where the measurement noise distributions among the reference nodes are all equal to find the optimal solution for the reference node deployment. By solving the gradient of the optimization objective function, we found that there is only one optimal deployment strategy according to the elevation angle, and the azimuth angle should be uniformly distributed. We then obtained a semi-closed form solution to obtain the optimal deployment.
	\item For large-scale network positioning, we considered more practical scenarios where multiple reference nodes locate multiple anchor nodes. In these scenarios, its hard to ensure that the measurement noise distribution among reference nodes are all the same for every anchor node. Therefore, we extended the situation to a more relaxed one where the measurement noise distributions among reference nodes are different. By analyzing the first-order derivative of the optimization objective function, we found that the optimal solution of reference node deployment is equivalent to the situation that the measurement noise distribution among reference nodes are all the same. This conclusion provides theoretical support for the selection of reference nodes.
	\item We analyzed the effect of reference node positioning errors on the CRLB and provided the upper and lower boundaries of the CRLB in extreme cases. By analyzing the projection of errors in the light of sight direction, we found that surface reference node positioning errors can sometimes reduce range measurement errors, while other times positioning errors increase range measurement errors.
\end{enumerate}

\indent The rest of the paper is organized as follows. The problem formulation is provided in Section \uppercase\expandafter{\romannumeral2}, and Section \uppercase\expandafter{\romannumeral3} gives the optimal deployment strategy of reference nodes. In section \uppercase\expandafter{\romannumeral4}, we did some simulations to verify the findings in section \uppercase\expandafter{\romannumeral3}. In section \uppercase\expandafter{\romannumeral5}, we conducted an ocean experiment to evaluate the localization performance with two different elevation angles. Finally, we draw a conclution in section \uppercase\expandafter{\romannumeral6}.

\section{System Model and Problem Formulation}
In this section, the system model for localizing seafloor anchor node is first presented, then the formulation of optimization problem is derived.
\subsection{System Model}
\begin{figure*}[htbp]
	\centering
	\includegraphics[width=0.7\linewidth]{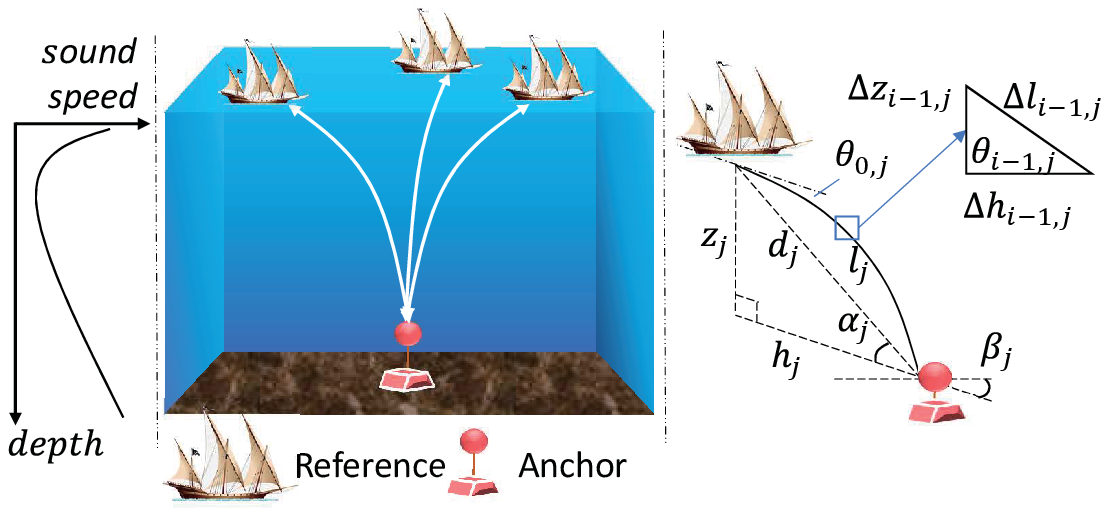}%
	\caption{Scenario of underwater target localization using J TOA anchor nodes. Between reference node j and the anchor, $z_j$ is the vertical distance, $h_j$ is the horizontal signal propagation distance, $d_j$ is the light of sight distance, $l_j$ is the real signal propagation distance, $\alpha_j$ is the elevation angle, $\beta_j$ is the azimuth angle, and $\theta_{0,j}$ is the initial grazing angle.}
	\label{fig01}
\end{figure*}
In order to locate seafloor anchor nodes, surface buoys can be used for positioning, or shipborne units can be used as virtual buoys for positioning, the later of which has lower economic costs but longer positioning cycles, and is adopted in this paper. 

\indent We consider using J TOA sonar reference points, whose locations are known by GPS with negligible errors, to localize a suspended underwater anchor node in the 3D space. Each node contains a transmitter and receiver, so that clock asynchronous errors can be eliminated through round--trip TOA process. The scenario of underwater target localization is shown in Fig.~\ref{fig01}. Due to the non-uniform distribution of sound speed, the signal will not propagate straightly with Snell effect.

\indent Let the sound speed profile (SSP) processed after standardized interpolation be $\boldsymbol{\mathcal{S}}_z=\{(s_0,z_0),(s_1,z_1)...(s_i,z_i)\},i=0,1,...,I$, in which $s_i$ is the sound speed value, $z_i$ is the depth value, and the depth is equally divided that $\Delta z_{i} = z_i-z_{i-1} = z_{i+1} - z_i = \Delta z_{i+1}$. For each linear depth layer $i$, there will be
\begin{equation}
	s_i = s_{i-1}(1+g_i),\label{eq1}
\end{equation}
where $g_i$ describes the gradient of sound speed variation.

According to the stratified ray tracing Model with constant sound speed in \cite{Wang2013UnderwaterAcoustics}, for each linear layer of the given SSP, the horizontal propagation distance can be calculated by
\begin{equation}
	\Delta h_{i,j} = \frac{\Delta z_{i}}{\tan\theta_{i-1,j}},\label{eq2}
\end{equation}
where $\theta_{i-1,j}$ is the grazing angle at the $i$th depth layer referring to reference node j. When the depth interval of the sound speed layer is small enough, the real propagation distance will be approximately equal to the line segment $\Delta l_{i,j}$, thus $\Delta l_{i,j}$ will be
\begin{equation}
	\Delta l_{i,j} = \frac{\Delta h_{i,j}}{cos\theta_{i-1,j}}=\frac{\Delta z_{i}}{\sin\theta_{i-1,j}},\label{eq3}
\end{equation}

\indent According to the Snell's law \cite{Jensen2011computational}
\begin{equation}
	\frac{cos\theta_{i,j}}{s_i}=\frac{cos\theta_{i-1,j}}{s_{i-1}}=\frac{cos\theta_{0,j}}{s_{0}},\label{eq4}
\end{equation}
where $s_{0}$ is the sound speed at the ocean surface, the total signal propagation path will be
\begin{equation}
	L_{j} = \sum_{i=1}^{I}\Delta l_{i,j} = \sum_{i=1}^{I}\frac{\Delta z_{i}}{\sin\theta_{i-1,j}},\label{eq5}
\end{equation}
which can be further derived as
\begin{equation}
	L_{j} =\sum_{i=1}^{I}\frac{s_0}{\sqrt{s_0^2 - s_{i-1}^2\cos^2\theta_{0,j}}},
\end{equation}
where it is assumed that $\Delta z_{i} = 1$ (standardized interpolation). And the total horizontal propagation distance can be calculated as
\begin{equation}
	h_{j} =\sum_{i=1}^{I}\frac{s_{i-1}\cos\theta_{0,j}}{\sqrt{s_0^2 - s_{i-1}^2\cos^2\theta_{0,j}}},
\end{equation}

Let $\Delta\hat{l}_{i,j} = \Delta l_{i,j} + n_{l,i,j}$, where $n_{l,i,j}\sim N\left(0,\sigma_{l,i,j}^2\right)$ is the measurement noise following Gaussian distribution, and $\sigma_{l,i,j} = \gamma\Delta l_{i,j}$ is a proportional function of the real signal propagation distance, where $\gamma$ is the proportional coefficient. Thus, there is $n_{l,i,j}\sim N\left(0,(\gamma\Delta l_{i,j})^2\right)\sim N\left(0,\frac{\gamma^2 s_0^2}{s_0^2 - s_{i-1}^2\cos^2\theta_{0,j}}\right)$. In the following lemma, we show that the equivalent light--of--sight measurement error still follows a Gaussian distribution.

\textit{\textbf{Lemma 1}}: The light--of--sight measurement error between reference node J and the target follows $n_{d,j}\sim N\left(0,\sigma^2_{d,j}\right)$ where $\sigma^2_{d,j} = \frac{cos^2\theta_{0,j}}{cos^2\alpha_{j}}\sum_{i=1}^I \frac{\gamma^2 s_{i-1}^2}{s_0^2 - s_{i-1}^2\cos^2\theta_{0,j}}$.

\textit{Proof}: Please refer to Appendix A. 

Without loss of generality, assume there are J reference nodes that are within the communication range of the target, the anchor node measurement covariance matrix will be
\begin{equation}
	\mathbf{\Sigma} = \begin{bmatrix}
		\sigma^2_{d,1}&&&\\
		&\sigma^2_{d,2}&&\\
		&&...&\\
		&&&\sigma^2_{d,J}\\
	\end{bmatrix}.
\end{equation}

If the depths of all reference nodes are all the same, and the measurement errors of signal propagation path follow the same distribution, then
the measurement covariance matrix will be
\begin{equation}
	\mathbf{\Sigma} = \begin{bmatrix}
		\sigma^2_{d,1}&&&\\
		&\sigma^2_{d,2}&&\\
		&&...&\\
		&&&\sigma^2_{d,J}\\
	\end{bmatrix} = \begin{bmatrix}
	\sigma^2_{d}&&&\\
	&\sigma^2_{d}&&\\
	&&...&\\
	&&&\sigma^2_{d}\\
	\end{bmatrix},
\end{equation}
where $\sigma^2_{d} = \frac{cos^2\theta_{0}}{cos^2\alpha}\sum_{i=1}^I \frac{\gamma^2 s_{i-1}^2}{s_0^2 - s_{i-1}^2\cos^2\theta_{0}}$, meaning the $\theta_{0,j}$ or $\alpha_{j}$ will be the same for different reference nodes.

\begin{figure}[!htbp]
	\centering
	\includegraphics[width=0.7\linewidth]{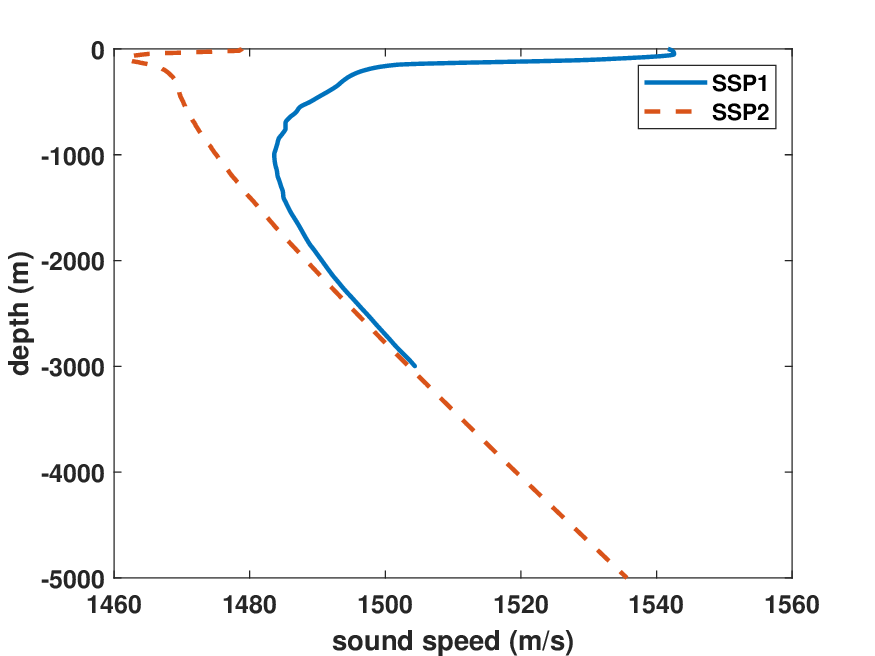}%
	\caption{Typical SSP distribution. SSP1 refers to typical SSP sampled at the southern Pacific Ocean, while SSP2 refers to typical SSP sampled at the northern Pacific Ocean.}
	\label{fig02}
\end{figure}
\begin{figure}[!htbp]
	\centering
	\includegraphics[width=0.7\linewidth]{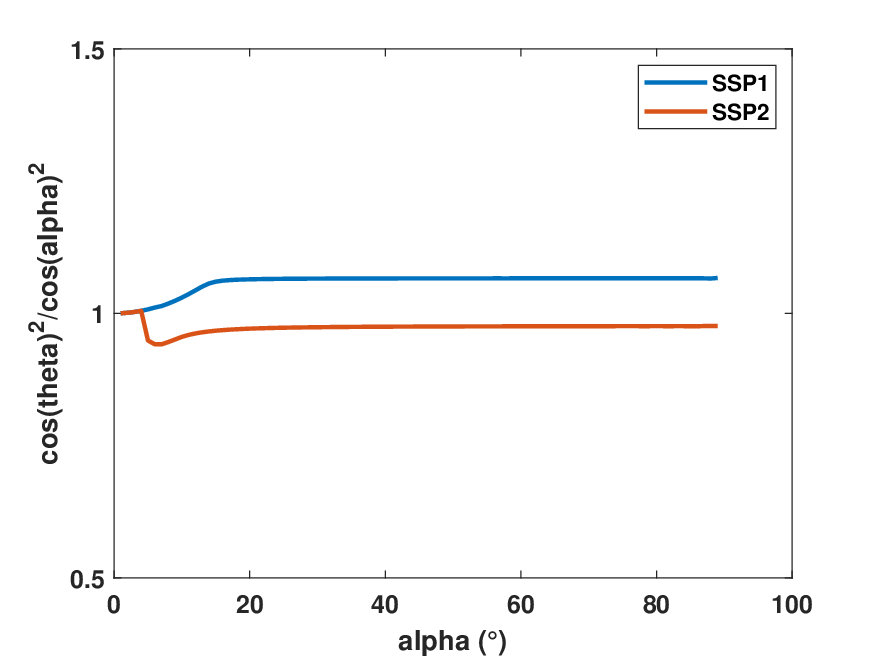}%
	\caption{Numerical relation test between $\alpha$ and $\theta$.}
	\label{fig03}
\end{figure}

Based on numerical tests with some typical SSP distributions, as shown in Fig.~\ref{fig02}, there is approximately $\cos^2\theta_0/\cos^2\alpha \approx 1$, which is shown in Fig.~\ref{fig03}. The inflection point of the curve represents the depth of the sound axis, and usually, the depth of the sound axis is deeper in low latitude areas and shallower in high latitude areas. As a result, it can be simplified that $\sigma^2_{d} \approx \sum_{i=1}^I \frac{\gamma^2 s_{i-1}^2}{s_0^2 - s_{i-1}^2\cos^2\alpha}$.

\subsection{Problem Formulation}
The Jacobian vector of the $j$th measurement error evaluated at the true target anchor node's position $\boldsymbol{p^t} = \left[p^{t}_{x}, p^{t}_{y}, p^{t}_{z}\right]^T$ can be expressed as
\begin{equation}
	\boldsymbol{\mathcal{J}}_{j} = \left.\frac{\partial \hat{d}_{j}}{\partial \mathbf{\textit{p}}^T}\right|_{\boldsymbol{p^t}} \\
	= \left.\left[\frac{\partial \hat{d}_{j}}{\partial p^{t}_{x}}, \frac{\partial \hat{d}_{j}}{\partial p^{t}_{y}}, \frac{\partial \hat{d}_{j}}{\partial p^{t}_{z}}\right]\right|_{\boldsymbol{p^t}}. \label{eq10}
\end{equation}

\textit{\textbf{Lemma 2}}: Let the position of the $j$th reference node be $\boldsymbol{p^{r}_{j}}$, and $ \boldsymbol{p^{r}_{j}} = \left[p^{r}_{x,j}, p^{r}_{y,j}, p^{r}_{z,j}\right]^T$, then the equation \eqref{eq10} can be derived as

\begin{equation}
	\boldsymbol{\mathcal{J}}_{j} = \left[\cos\alpha_j \cos\beta_j \quad \cos\alpha_j \sin\beta_j \quad \sin\alpha_j\right]. \label{eq11}
\end{equation}

\textit{Proof}: Please refer to Appendix B. 

Based on equation \eqref{eq11}, the Jacobian matrix of the J measurements can be obtained as
\begin{equation}
	\boldsymbol{\mathcal{J}}_0 = \begin{bmatrix}
		\cos\alpha_1 \cos\beta_1& \cos\alpha_1 \sin\beta_1 & \sin \alpha_1\\
		\cos\alpha_2 \cos\beta_2& \cos\alpha_2 \sin\beta_2 & \sin \alpha_2\\
		... &...& ...\\
		\cos\alpha_J \cos\beta_J& \cos\alpha_J \sin\beta_J & \sin \alpha_J\\
	\end{bmatrix}_{J \times 3}.
\end{equation}

Each column element of $\boldsymbol{\mathcal{J}}_0$ has the same form of expression, so it can be simplified by using vectors. For simplification reason, we define

\begin{equation}
	\begin{split}
		\boldsymbol{a} &= \left[\cos\alpha_1 \cos\beta_1,...,\cos\alpha_J \cos\beta_J\right]^T\\
		\boldsymbol{b} &= \left[\cos\alpha_1 \sin\beta_1,...,\cos\alpha_J \sin\beta_J\right]^T\\
		\boldsymbol{c} &= \left[\sin\alpha_1,...,\sin\alpha_J\right]^T\\
		\boldsymbol{\mathcal{J}}_0 & = \left[\boldsymbol{a},\boldsymbol{b},\boldsymbol{c}\right]_{J \times 3}.
	\end{split}
\end{equation}

To obtain the CRLB, the FIM needs to be calculated. The FIM $\boldsymbol{\Phi}$ is given by \cite{Torrieri1984Location} as

\begin{equation}
	\begin{split}
	\boldsymbol{\Phi} &= \left[\boldsymbol{a} \quad \boldsymbol{b} \quad \boldsymbol{c}\right]^T \mathbf{\Sigma}^{-1}\left[\boldsymbol{a} \quad \boldsymbol{b} \quad \boldsymbol{c}\right]\\
	&=\begin{bmatrix}
		\boldsymbol{a}^T\mathbf{\Sigma}^{-1}\boldsymbol{a}& \boldsymbol{a}^T\mathbf{\Sigma}^{-1}\boldsymbol{b} & \boldsymbol{a}^T\mathbf{\Sigma}^{-1}\boldsymbol{c}\\
		\boldsymbol{b}^T\mathbf{\Sigma}^{-1}\boldsymbol{a}& \boldsymbol{b}^T\mathbf{\Sigma}^{-1}\boldsymbol{b} & \boldsymbol{b}^T\mathbf{\Sigma}^{-1}\boldsymbol{c}\\
		\boldsymbol{c}^T\mathbf{\Sigma}^{-1}\boldsymbol{a}& \boldsymbol{c}^T\mathbf{\Sigma}^{-1}\boldsymbol{b} & \boldsymbol{c}^T\mathbf{\Sigma}^{-1}\boldsymbol{c}\\
	\end{bmatrix}_{3 \times 3}.
	\end{split}
\end{equation}
 Let $\boldsymbol{\hat{a}} = \mathbf{\Sigma}^{-1/2}\boldsymbol{a}$, $\boldsymbol{\hat{b}} = \mathbf{\Sigma}^{-1/2}\boldsymbol{b}$, $\boldsymbol{\hat{c}} = \mathbf{\Sigma}^{-1/2}\boldsymbol{c}$, there is
\begin{equation}
 	\begin{split}
 		\boldsymbol{\Phi} =\begin{bmatrix}
 			\boldsymbol{\hat{a}}^T\boldsymbol{\hat{a}}& \boldsymbol{\hat{a}}^T\boldsymbol{\hat{b}} & \boldsymbol{\hat{a}}^T\boldsymbol{\hat{b}}\\
 			\boldsymbol{\hat{b}}^T\boldsymbol{\hat{a}}& \boldsymbol{\hat{b}}^T\boldsymbol{\hat{b}} & \boldsymbol{\hat{b}}^T\boldsymbol{\hat{b}}\\
 			\boldsymbol{\hat{c}}^T\boldsymbol{\hat{a}}& \boldsymbol{\hat{c}}^T\boldsymbol{\hat{b}} & \boldsymbol{\hat{c}}^T\boldsymbol{\hat{c}}\\
 		\end{bmatrix}_{3 \times 3}.
 	\end{split}
\end{equation}

According to the A--optimality criterion that minimizing the trace of the inverse FIM in \cite{Xu2017OptimalAOA}, the CRLB will be

\begin{equation}
	CRLB = \boldsymbol{\Phi}^{-1}.
\end{equation}

\section{Optimization Deployment of Reference Nodes}
In this section, the optimization problem will first be simplified, then the same or different measurement noise among reference nodes will be discussed and optimal solution will be given according to the simplified problem, finally we will discuss the influence of position errors of reference nodes on the optimal solution.
\subsection{Optimization Problem Simplification}
Let the angle difference between vector $\boldsymbol{\hat{a}}$ and $\boldsymbol{\hat{b}}$ be $\phi_1$, between $\boldsymbol{\hat{a}}$ and $\boldsymbol{\hat{c}}$ be $\phi_2$, and between $\boldsymbol{\hat{b}}$ and $\boldsymbol{\hat{c}}$ be $\phi_3$, we have equation \eqref{eq17}.

\setcounter{equation}{16}
\begin{figure*}
	\begin{equation}
		\boldsymbol{\Phi}^{-1} = \frac{1}{\left|\boldsymbol{\Phi}\right|}\begin{bmatrix}
			|\boldsymbol{\hat{b}}|^2|\boldsymbol{\hat{c}}|^2(1-\cos^2\phi_3) & |\boldsymbol{\hat{a}}||\boldsymbol{\hat{b}}||\boldsymbol{\hat{c}}|^2(\cos\phi_2\cos\phi_3-\cos\phi_1) & |\boldsymbol{\hat{a}}||\boldsymbol{\hat{b}}|^2|\boldsymbol{\hat{c}}|(\cos\phi_1\cos\phi_3-\cos\phi_2)\\
			|\boldsymbol{\hat{a}}||\boldsymbol{\hat{b}}||\boldsymbol{\hat{c}}|^2(\cos\phi_2\cos\phi_3-\cos\phi_1)& |\boldsymbol{\hat{a}}|^2|\boldsymbol{\hat{c}}|^2(1-\cos^2\phi_2) & |\boldsymbol{\hat{a}}|^2|\boldsymbol{\hat{b}}||\boldsymbol{\hat{c}}|(\cos\phi_1\cos\phi_2-\cos\phi_3)\\
			|\boldsymbol{\hat{a}}||\boldsymbol{\hat{b}}|^2|\boldsymbol{\hat{c}}|(\cos\phi_1\cos\phi_3-\cos\phi_2)& |\boldsymbol{\hat{a}}|^2|\boldsymbol{\hat{b}}||\boldsymbol{\hat{c}}|(\cos\phi_1\cos\phi_2-\cos\phi_3)& |\boldsymbol{\hat{a}}|^2|\boldsymbol{\hat{b}}|^2(1-\cos^2\phi_1)\\
		\end{bmatrix}.\label{eq17}
	\end{equation}
\end{figure*}

\setcounter{equation}{29}
\begin{figure*}[!htbp]
	\begin{equation}
		\begin{split}
			&\frac{\partial \text{tr}(CRLB)}{\partial \alpha} = \left[ \frac{4J\gamma^2\sum_{i=1}^{I}M_i(\alpha)-4J\gamma^2\cos^2\alpha\sum_{i=1}^{I}(M_i(\alpha))^2}{(J\cos^2\alpha)^2}-\frac{J\gamma^2\sum_{i=1}^{I}M_i(\alpha)+J\gamma^2\sin^2\alpha\sum_{i=1}^{I}(M_i(\alpha))^2}{(J\sin^2\alpha)^2} \right]\sin2\alpha.
		\end{split}\label{eq30}
	\end{equation}
\end{figure*}

The detailed expression of $\left|\boldsymbol{\Phi}\right|$ is calculated by
\setcounter{equation}{17}
\begin{equation}
	\left|\boldsymbol{\Phi}\right| = |\boldsymbol{\hat{a}}|^2|\boldsymbol{\hat{b}}|^2|\boldsymbol{\hat{c}}|^2\lambda,\label{eq18}
\end{equation}
where $\lambda = (1-\cos^2\phi_1-\cos^2\phi_2-\cos^2\phi_3+2\cos\phi_1$ $\cos\phi_2\cos\phi_3)$.
Then, the objective function of the optimization problem, the trace of CRLB, becomes

\begin{equation}
	\begin{split}
		\text{tr}(CRLB) &= \frac{|\boldsymbol{\hat{b}}|^2|\boldsymbol{\hat{c}}|^2(1-\cos^2\phi_3)}{\left|\boldsymbol{\Phi}\right|} \\
		&+ \frac{|\boldsymbol{\hat{a}}|^2|\boldsymbol{\hat{c}}|^2(1-\cos^2\phi_2)}{\left|\boldsymbol{\Phi}\right|} \\
		&+ \frac{|\boldsymbol{\hat{a}}|^2|\boldsymbol{\hat{b}}|^2(1-\cos^2\phi_1)}{\left|\boldsymbol{\Phi}\right|} \\
		&= \frac{(1-\cos^2\phi_3)}{|\boldsymbol{\hat{a}}|^2\lambda} +\frac{(1-\cos^2\phi_2)}{|\boldsymbol{\hat{b}}|^2\lambda} +\frac{(1-\cos^2\phi_1)}{|\boldsymbol{\hat{c}}|^2\lambda}.  \\
	\end{split}\label{eq19}
\end{equation}

Accordingly, we establish the following proposation to simplify the optimization problem.

\textit{\textbf{Proposation 1}}: Based on equation \eqref{eq19}, there is

\begin{equation}
	\begin{split}
		\text{tr}(CRLB)
		&\geq \frac{1}{|\boldsymbol{\hat{a}}|^2} + \frac{1}{|\boldsymbol{\hat{b}}|^2} + \frac{1}{|\boldsymbol{\hat{c}}|^2},
	\end{split}\label{eq20}
\end{equation}

with equality if 
\begin{equation}
	\cos\phi_1 =\cos\phi_2 = \cos\phi_3=0.
\end{equation}

\textit{Proof}: Please refer to Appendix C.

Actually there is relationship that

\begin{equation}
	|\boldsymbol{\hat{a}}|^2 = \sum_{j=1}^{J} \frac{\cos^2 \alpha_j \cos^2 \beta_j}{\sigma_d^2}
\end{equation}
\begin{equation}
	|\boldsymbol{\hat{b}}|^2 = \sum_{j=1}^{J} \frac{\cos^2 \alpha_j \sin^2 \beta_j}{\sigma_d^2}
\end{equation}
\begin{equation}
	|\boldsymbol{\hat{c}}|^2 = \sum_{j=1}^{J} \frac{\sin^2 \alpha_j}{\sigma_d^2}.
\end{equation}

The minimizing problem becomes
\begin{equation}
	\begin{split}
		\arg\min\limits_{\{\alpha_j,\beta_j\},j=1,...,J} \left[\left(\sum_{j=1}^{J} \frac{\cos^2 \alpha_j \cos^2 \beta_j}{\sigma_d^2}\right)^{-1}\right.\\
		\left.+ \left(\sum_{j=1}^{J} \frac{\cos^2 \alpha_j \sin^2 \beta_j}{\sigma_d^2}\right)^{-1} + \left(\sum_{j=1}^{J} \frac{\sin^2 \alpha_j}{\sigma_d^2}\right)^{-1}\right],\\
	\end{split}\label{eq25}
\end{equation}
subject to
\begin{equation}
	\begin{split}
		\sum_{j=1}^{J} \frac{\cos^2 \alpha_j \cos\beta_j \sin\beta_j}{\sigma_d^2} = 0\\
		\sum_{j=1}^{J} \frac{\cos\alpha_j \cos\beta_j \sin\alpha_j}{\sigma_d^2} = 0\\
		\sum_{j=1}^{J} \frac{\cos\alpha_j \sin\beta_j \sin\alpha_j}{\sigma_d^2} = 0.
	\end{split}\label{eq26}
\end{equation}

This is a constrained nonlinear optimization problem, which can be solved by the method of Lagrange multipliers. However, it is quite complicated and we will use some simpler methods to solve it as discussed later.

\textit{\textbf{Proposation 2}}: The minimizing problem \eqref{eq25} can be further derived as
\begin{equation}
	\text{tr}(CRLB)	\geq \frac{4\sigma_d^2}{\sum_{j=1}^{J}\cos^2 \alpha_j} + \frac{\sigma_d^2}{\sum_{j=1}^{J}\sin^2 \alpha_j},\label{eq27}
\end{equation}
subject to
\begin{equation}
	\begin{split}
		\sum_{j=1}^{J} \cos^2 \alpha_j \cos\beta_j \sin\beta_j = \sum_{j=1}^{J} \cos\alpha_j \cos\beta_j \sin\alpha_j\\
		= \sum_{j=1}^{J} \cos\alpha_j \sin\beta_j \sin\alpha_j = \sum_{j=1}^{J} \cos^2 \alpha_j \cos2\beta_j = 0. \\
	\end{split}\label{eq28}
\end{equation}
\textit{Proof}: Please refer to Appendix D.

\subsection{Same measurement noise distribution among reference nodes}
\indent When $\sigma_{d,1} = \sigma_{d,2} = ... =\sigma_{d,J}$, it means that $\alpha_{1} = \alpha_{2} = ... = \alpha_{J}$ because the real propagation distance and horizontal propagation distance are all monotonic functions of the initial grazing angle \cite{Huang2024Localization}. Therefore, \eqref{eq27} can be derived as:

\begin{equation}
	\text{tr}(CRLB)	\geq \frac{4\sigma_d^2}{J\cos^2 \alpha} + \frac{\sigma_d^2}{J\sin^2 \alpha},\label{eq29}
\end{equation}

To find the optimal solution, we make partial derivative of \eqref{eq29} respect to $\alpha$, which can be computed as \eqref{eq30}
where $M_i(\alpha) = \frac{s_{i-1}^2}{s_0^2-s_{i-1}^2\cos^2\alpha}$. To further obtain the optimal solution, we make $\frac{\partial \text{tr}(CRLB)}{\partial \alpha}=0$ so that
\setcounter{equation}{30}
\begin{equation}
	\sin2\alpha=0,\label{eq31}
\end{equation}
or
\begin{equation}
	\begin{split}
	\frac{4J\gamma^2\sum_{i=1}^{I}M_i(\alpha)-4J\gamma^2\cos^2\alpha\sum_{i=1}^{I}(M_i(\alpha))^2}{(J\cos^2\alpha)^2}\\
	=\frac{J\gamma^2\sum_{i=1}^{I}M_i(\alpha)+J\gamma^2\sin^2\alpha\sum_{i=1}^{I}(M_i(\alpha))^2}{(J\sin^2\alpha)^2}.\label{eq32}
	\end{split}
\end{equation}

Two different possible optimal results can be obtained by case 1) substituting \eqref{eq31} into \eqref{eq29}, and case 2) substituting \eqref{eq32} into \eqref{eq29}. To satisfy condition $\sin2\alpha=0$, there is $\alpha\in\left\{0,\pi/2,-\pi/2\right\}$. As the reference nodes are on the ocean surface, and the anchor node is deployed at the bottom, the only case to satisfy $\sin2\alpha=0$ is $\alpha_1=\alpha_2=...=\alpha_J=\pi/2$. At this time, all reference nodes overlap into one point, which cannot meet the basic requirements of the spherical intersection positioning model. Therefore, case 1) cannot obtain the optimal solution we want.

\indent For case 2), it is hard to find an optimal closed form solution of $\alpha$, however, in the following corollary, we will prove that for the same measurement noise distribution among reference nodes, there exist and only one $\alpha$ that satisfies case 2), so that it will be possible to find feasible solutions through searching methods.

\textit{\textbf{Corollary 1}}: For the same measurement noise distribution among reference nodes, there exist and only one $\alpha$ that satisfies equation \eqref{eq32}.

\textit{Proof}: Please refer to Appendix E.

\indent Then, to satisfy \eqref{eq28}, we find two different groups of solutions, i.e. 1) J reference nodes consist one set, 2) J reference nodes are separate into multiple sub-sets.

\indent 1) If $J$ reference nodes consist one set:
\begin{equation}
	\beta_j = \frac{360^\circ(j-1)}{J} + \beta_0,\label{eq33}
\end{equation}
where $\beta_0$ can be any value. In fact, \eqref{eq33} means that $\beta_j$ and $2\beta_j$ of reference nodes should have equal angular distribution.

\indent 2) If $J$ reference nodes are separated into $Q$ multiple subsets with $J_q\geq3,q=1,2,...,Q$ reference nodes for each sub-set:
\begin{equation}
	\beta_j = \frac{360^\circ(j-1)}{J_q} + \beta_0.\label{eq34}
\end{equation}
There is no geometrical requirement between difference sub-sets. However, when there are enough reference nodes, it is easy to form a circular reference node array.

\subsection{Different measurement noise distribution among reference nodes}
As an extension of the above analysis, the optimal reference deployment strategy under a more practical circumstance that difference nodes have different signal measurement noise variance $\sigma_{d,j}$ is developed in this section.

With different $\sigma_{d,j}$, the optimal problem \eqref{eq25} will be:
\begin{equation}
	\text{tr}(CRLB)	\geq \sum_{j=1}^{J}\frac{4\sigma_{d,j}^2}{\cos^2 \alpha_j} + \sum_{j=1}^{J}\frac{\sigma_{d,j}^2}{\sin^2 \alpha_j},\label{eq35}
\end{equation}
the equality is reached when
\begin{equation}
\sum_{j=1}^{J} \frac{\cos^2 \alpha_j \cos^2 \beta_j}{\sigma_{d,j}^2} = \sum_{j=1}^{J} \frac{\cos^2 \alpha_j \sin^2 \beta_j}{\sigma_{d,j}^2},
\end{equation}
and condition \eqref{eq26} is satisfied, which is
\begin{equation}
	\begin{split}
		\sum_{j=1}^{J} \frac{\cos^2 \alpha_j \sin2\beta_j}{\sigma_{d,j}^2} = \sum_{j=1}^{J} \frac{\cos^2 \alpha_j \cos2\beta_j}{\sigma_{d,j}^2}= 0\\
		\sum_{j=1}^{J} \frac{\sin2\alpha_j\cos\beta_j }{\sigma_{d,j}^2} = \sum_{j=1}^{J} \frac{\sin2\alpha_j\sin\beta_j }{\sigma_{d,j}^2} = 0.
	\end{split}
\end{equation}

According to Fig.~\ref{fig03}, there is $\sigma^2_{d,j} \approx \sum_{i=1}^I \frac{\gamma^2 s_{i-1}^2}{s_0^2 - s_{i-1}^2\cos^2\alpha_{j}}$. To find the optimal solution of $\alpha_j$, we make partial derivative of \eqref{eq35} respect to $\alpha_j$, which can be computed as \eqref{eq38} with $M_i(\alpha_j) = \frac{s_{i-1}^2}{s_0^2-s_{i-1}^2\cos^2\alpha_j}$. In the following corollary, we will prove that there exists and only one optimal solution for different measurement noise distribution among reference nodes.

\begin{figure*}
	\begin{equation}
		\begin{split}
			&\frac{\partial \text{tr}(CRLB)}{\partial \alpha_j} = \left[ \frac{4\gamma^2\sum_{i=1}^{I}M_i(\alpha_j)-4\gamma^2\cos^2\alpha_j\sum_{i=1}^{I}(M_i(\alpha_j))^2}{(\cos^2\alpha_j)^2}-\frac{\gamma^2\sum_{i=1}^{I}M_i(\alpha_j)+\gamma^2\sin^2\alpha_j\sum_{i=1}^{I}(M_i(\alpha_j))^2}{(\sin^2\alpha_j)^2} \right]\sin2\alpha_j,
		\end{split}\label{eq38}
	\end{equation}
\end{figure*}

\textit{\textbf{Corollary 2}}: $\alpha_1 = \alpha_2 = ... = \alpha_J = a$ is and the only optimal solution for different measurement noise distribution among reference nodes.

\textit{Proof}: Please refer to Appendix F.

Corollary 2 indicates that it is actually equivalent to the situation where the error of each node is the same.

\subsection{Influence of reference node positioning error}
Although the position of the surface reference node is obtained through GPS, positioning errors may still occur due to factors such as ocean waves affecting the attitude of the surface reference nodes. In this section, we are going to analyze the influence of reference nodes' positioning errors on the trace of CRLB.

\indent Within the vertical profile between the reference node and the target node, assume the position error of reference node $j$ is $\Delta\boldsymbol{p^r_j}$ and follows Gaussian distribution as $\Delta\boldsymbol{p^r_j} \sim N(0,\sigma_{r,j}^2)$, so the actual position of reference node $j$ will be $\boldsymbol{\hat{p}^r_j} = \boldsymbol{p^r_j} + \Delta\boldsymbol{p^r_j}$. Then the projection of positioning error on the oblique distance between the reference node and the target node is $\Delta\boldsymbol{p^r_j} \cos \alpha_j$, therefore variance of the measurement noise will be $\hat{\sigma}_d^2 = \sigma_d^2 \pm \sigma_{r,j}^2\cos^2\alpha_j$. Assume the position error among all reference nodes follow the same distribution, the optimization problem of equation \eqref{eq29} becomes 

\begin{equation}
	\text{tr}(CRLB)	\geq \frac{4(\sigma_d^2 \pm \sigma_{r}^2\cos^2\alpha)}{J\cos^2 \alpha} + \frac{(\sigma_d^2 \pm \sigma_{r}^2\cos^2\alpha)}{J\sin^2 \alpha}, \label{eq39}
\end{equation}
and equation \eqref{eq28} is satisfied. The equation \eqref{eq39} means that sometimes, positioning errors of surface reference nodes may reduce the ranging measurement errors, while other times positioning errors increase ranging errors. The lower bound of CRLB is

\begin{equation}
	\text{tr}(CRLB)_{low}	= \frac{4(\sigma_d^2 - \sigma_{r}^2\cos^2\alpha)}{J\cos^2 \alpha} + \frac{(\sigma_d^2 - \sigma_{r}^2\cos^2\alpha)}{J\sin^2 \alpha}, \label{eq40}
\end{equation}
and the upper bound of CRLB is

\begin{equation}
	\text{tr}(CRLB)_{up}	= \frac{4(\sigma_d^2 + \sigma_{r}^2\cos^2\alpha)}{J\cos^2 \alpha} + \frac{(\sigma_d^2 + \sigma_{r}^2\cos^2\alpha)}{J\sin^2 \alpha}. \label{eq41}
\end{equation}

\section{Simulation Results}
\indent In this section, we will conduct several simulations to demonstrate the findings in section \uppercase\expandafter{\romannumeral2} and \uppercase\expandafter{\romannumeral3}. 

\subsection{Existence of optimal solution}
\begin{figure}[!htbp]
	\centering
	\includegraphics[width=0.7\linewidth]{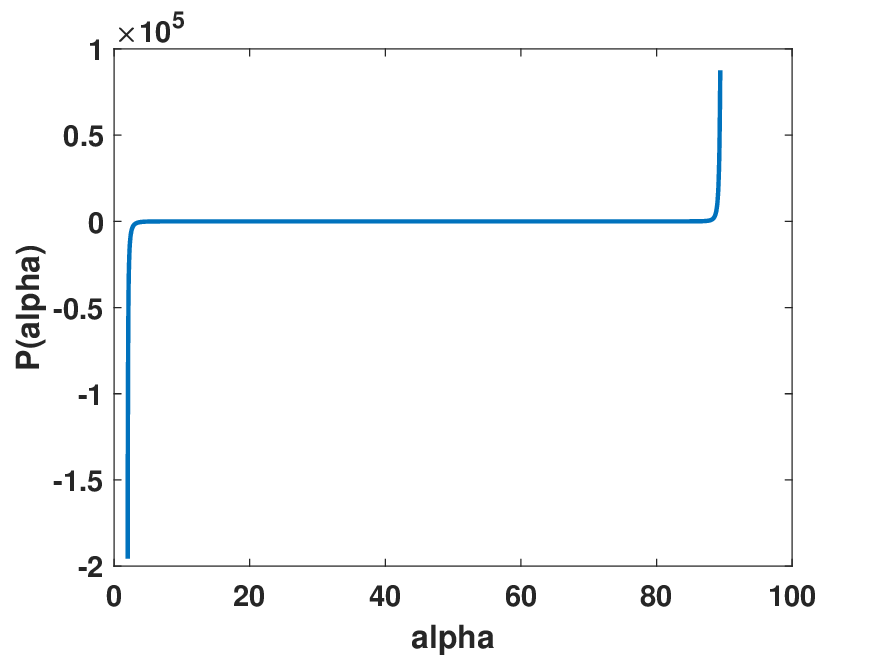}%
	\caption{$P(\alpha)$ as a function of angle $\alpha$.}
	\label{fig04}
\end{figure}

\begin{figure}[!htbp]
	\centering
	\includegraphics[width=0.7\linewidth]{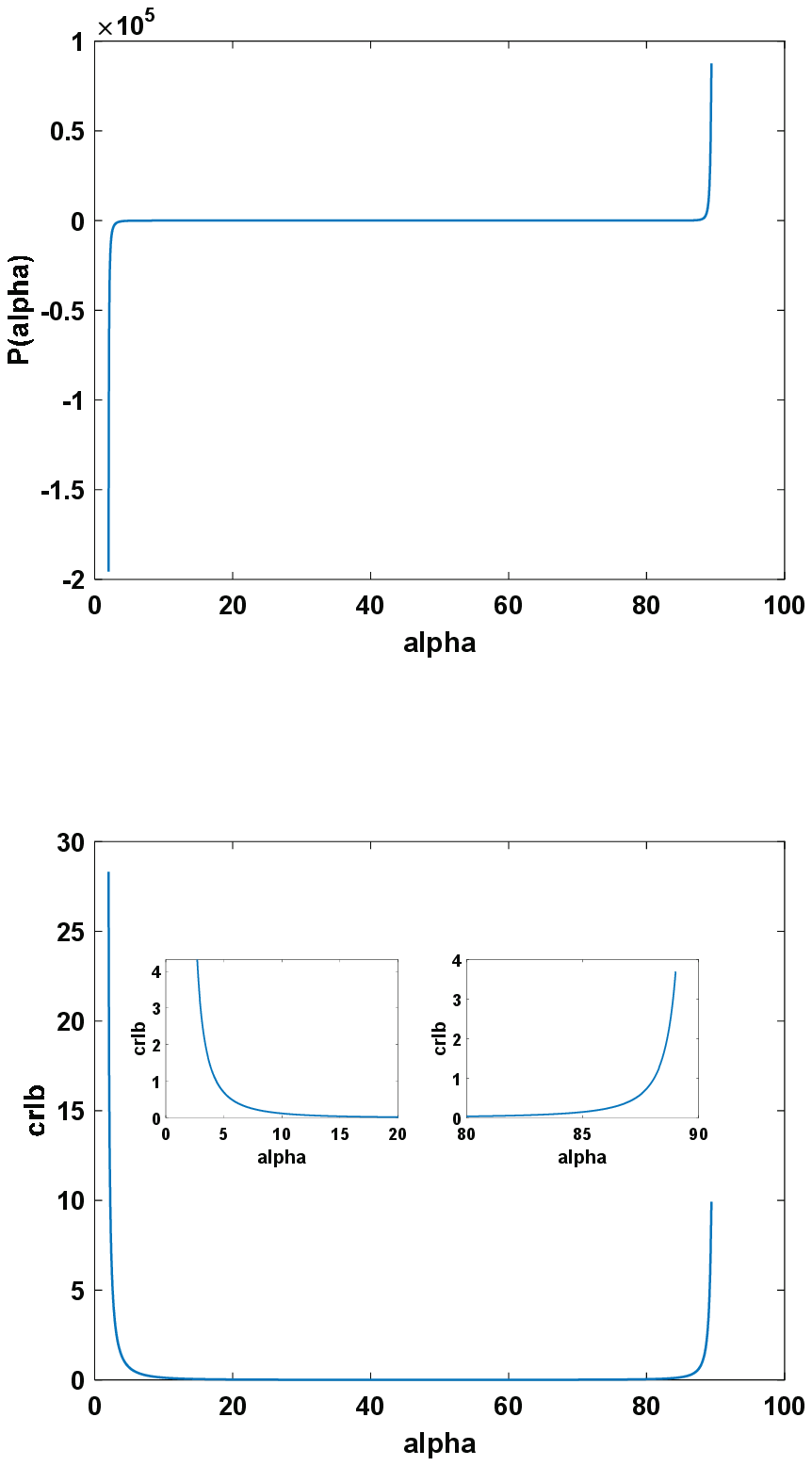}%
	\caption{$tr(CRLB)$ as a function of angle $\alpha$.}
	\label{fig05}
\end{figure}

First of all, to verify the existence and uniqueness of the optimal solution for the elevation angle. The $P(\alpha)$ in Appendix E and $tr(CRLB)$ as functions of angle $\alpha$ are separately given in Fig.~\ref{fig04} and Fig.~\ref{fig05} based on the solid line of SSP in Fig.~\ref{fig02}. It shows that $P(\alpha)$ is an increasing function from negative infinity to positive infinity, which goes through the stages of rapid growth, stationary growth, and second rapid growth. Therefore $P(\alpha)=0$ has a unique solution. In Fig.~\ref{fig05}, the curve of $tr(CRLB)$ is a convex function and has a minimum value according to corollary 1, which is 0.0052 when $\alpha = 48.4^\circ$.

\subsection{Localization error as a function of the elevation angle}

\indent Since we are looking for the optimal reference node deployment scheme for underwater localization of a given anchor node, the most direct performance verification indicator is the positioning accuracy of the target. There are many localization algorithms for underwater target positioning, and we simply adopt the iterative depth fine-tuning positioning algorithm proposed in our previous work, which can be detaily found in \cite{Huang2024Localization}. The parameter settings of the localization algorithm is shown in Table~\ref{table1}, where the initial direction of depth fine-tuning is towards the decreasing of depth with value of 1.

\begin{table}[!htbp]
	\caption{Parameter settings of localization algorithm \cite{Huang2024Localization}\label{table1}}
	\centering
	\begin{tabular}{|c||c|}
		\hline
		number of reference nodes & 5\\
		\hline
		number of anchor node & 1\\
		\hline
		number of tests & 50\\
		\hline
		threshold of depth step & 0.2 (m)\\
		\hline
		initial direction of depth fine-tuning & 1\\
		\hline
		initial depth fine-tuning step & 2 (m)\\
		\hline
		start depth & 2505 (m)\\
		\hline
	\end{tabular}
\end{table}

\begin{figure}[!htbp]
	\centering
	\includegraphics[width=0.8\linewidth]{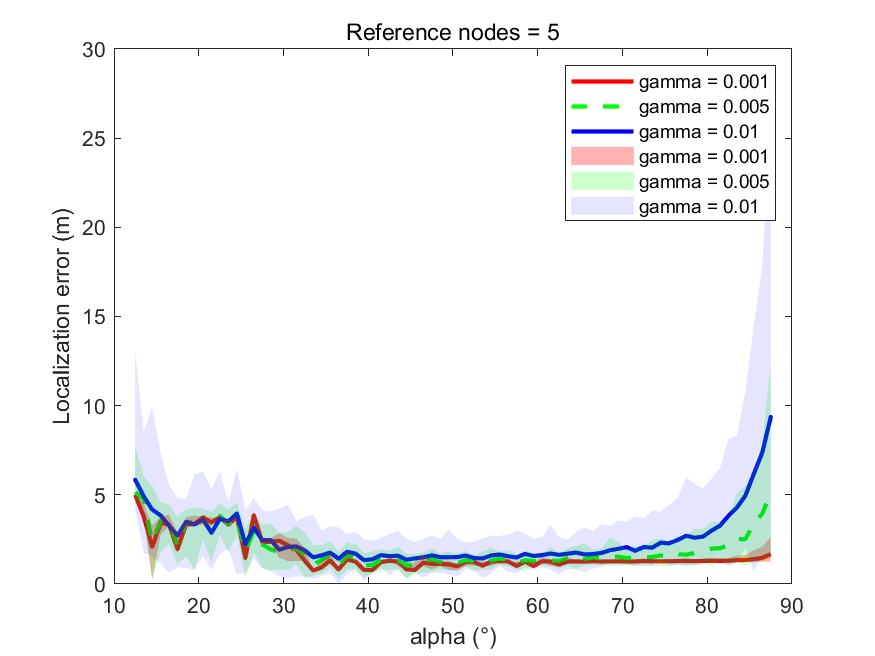}%
	\caption{Localization error with the variation of $\alpha$.}
	\label{fig06}
\end{figure}

\begin{figure*}[!htbp]
	\centering
	\subfloat[]{
		\includegraphics[width=0.3\linewidth]{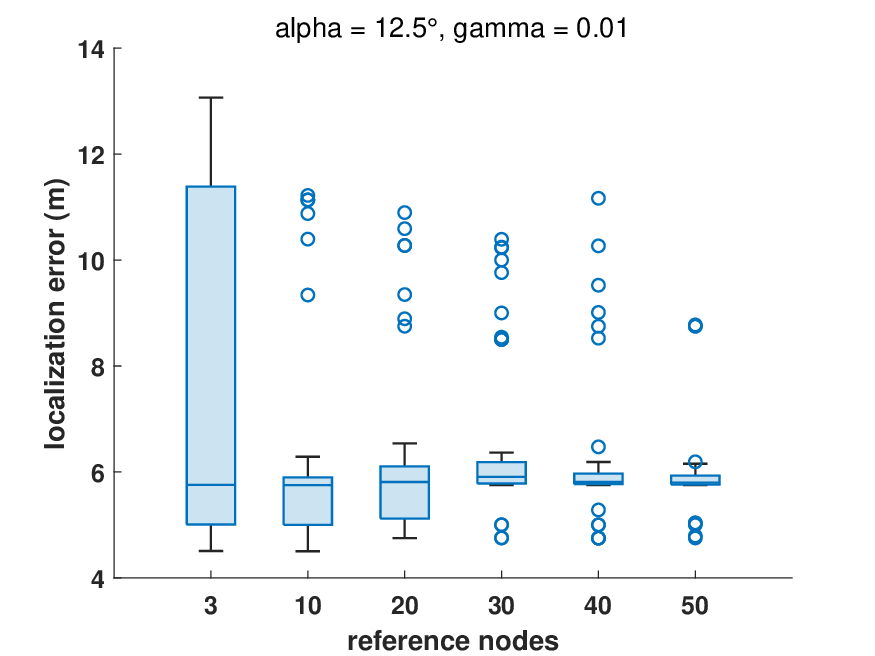}\label{fig07a}%
	}
	\subfloat[]{
		\includegraphics[width=0.3\linewidth]{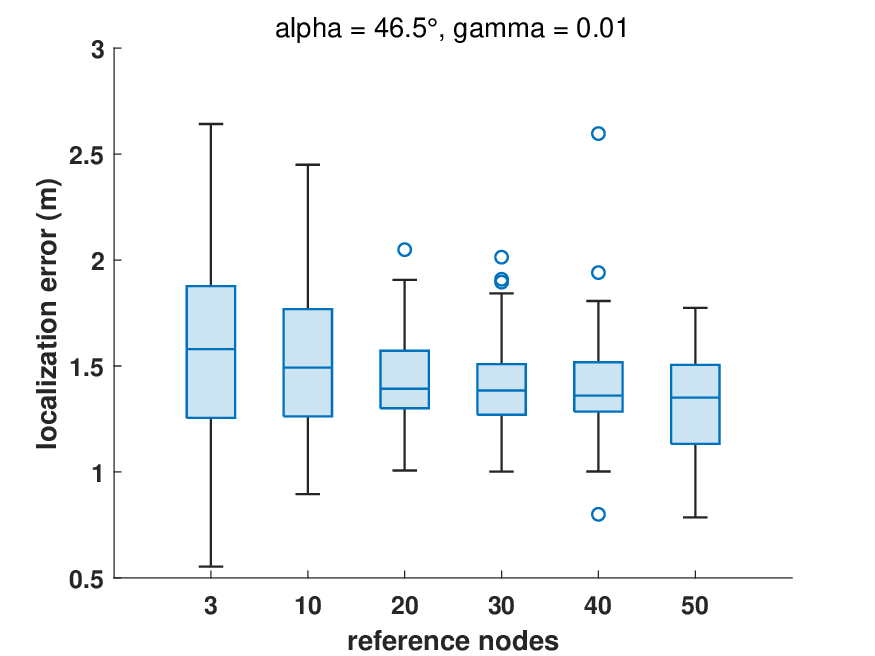}\label{fig07b}%
	}
	\subfloat[]{
		\includegraphics[width=0.3\linewidth]{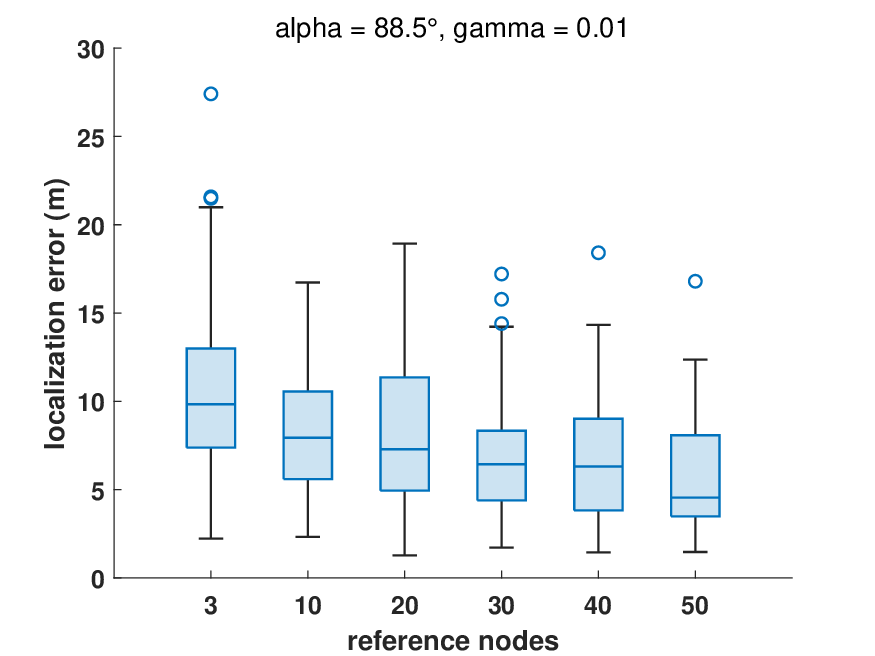}\label{fig07c}%
	}\\
	\subfloat[]{
		\includegraphics[width=0.3\linewidth]{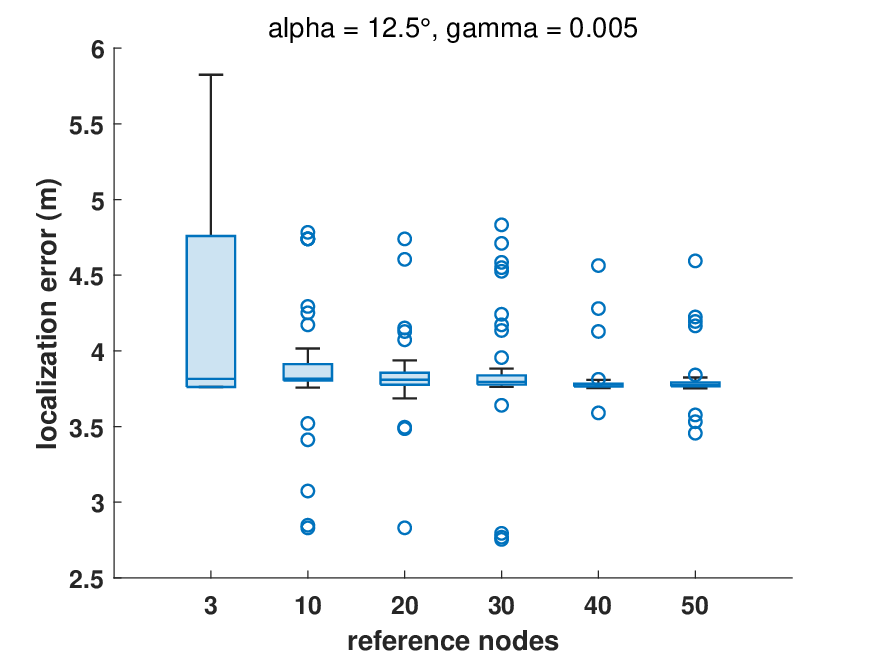}\label{fig07d}%
	}
	\subfloat[]{
		\includegraphics[width=0.3\linewidth]{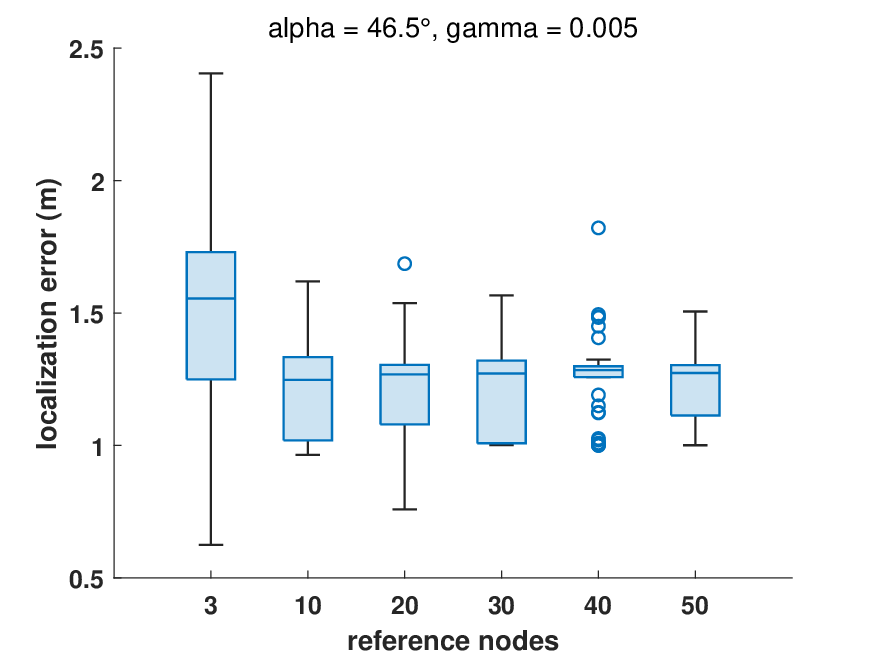}\label{fig07e}%
	}
	\subfloat[]{
		\includegraphics[width=0.3\linewidth]{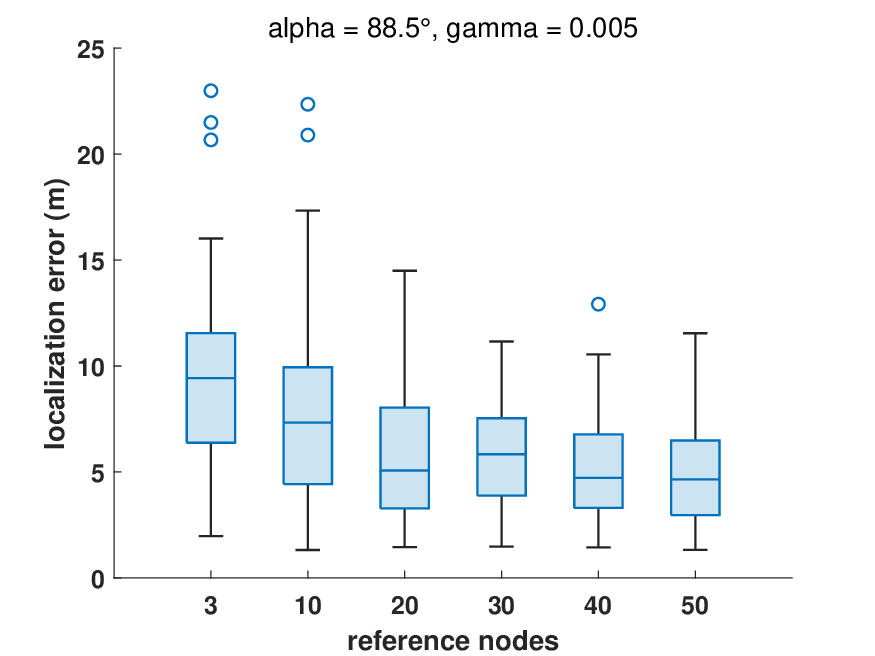}\label{fig07f}%
	}
	\caption{Localization error with different number of reference nodes.}
	\label{fig07}
\end{figure*}

To test the localization errors with different elevation angles, we conduct a simulation that 5 surface reference nodes are uniformly distributed in a circular shape on the horizontal plane, and these nodes collectively locate a anchor node deployed on the seabed. From a top view perspective, the anchor node is located at the center of the circle. The localization errors under 50 test times are shown in Fig~\ref{fig06} with different gamma parameter. As the noise level ($\gamma$) increases, the average localization error grows and the range of error variation becomes more unstable. This phenomenon is constent with conventional cognition. For a given noise level, such as $\gamma = 0.01$, the localization error first decreases to an minimum value and the grows, which has the same changing trend with $tr(CRLB)$, indicating the accuracy of the derived $tr(CRLB)$. The optimal evelation angle solution according to different $\gamma$ is given in Table~\ref{table2}. There is a slight deviation between the optimal elevation angle of positioning simulation and the theoretical analysis of $48.4^\circ$, mainly due to the randomness of positioning testing and added noises. However, the positioning error curve tends to stabilize around $48.4^\circ$, so the deviation between simulation and theory is within an acceptable range.

\begin{table}[!htbp]
	\caption{Optimal solution of evelation angle\label{table2}}
	\centering
	\begin{tabular}{|c||c|}
		\hline
		$\gamma$ & optimal simulation elevation angle $\alpha^\circ$\\
		\hline
		0.01 &  45.5\\
		\hline
		0.005 & 45.5\\
		\hline
		0.001 & 46.5\\
		\hline
	\end{tabular}
\end{table}

\subsection{Influence of reference node number}

To test the influence of the number of reference nodes, the localization error of anchor node as a function of reference node numbers is given in Fig~\ref{fig07}, where Fig~\ref{fig07a}, Fig~\ref{fig07b}, Fig~\ref{fig07c} correspond to $\gamma = 0.01$ and Fig~\ref{fig07d}, Fig~\ref{fig07e}, Fig~\ref{fig07f} correspond to $\gamma = 0.005$. All reference nodes are uniformly distributed to form a circle form in this simulation. It is obvious that the localization error when $\alpha = 46.5$ is much smaller than the localization error when $\alpha = 12.5$ and $\alpha = 88.5$, which is consistent with that shown in Fig~\ref{fig06}. From Fig~\ref{fig07}, the trend of localization error is gradually decrease as the number of reference nodes increase. When the number of reference nodes exceeds 10, the overall change tends to stabilize and there is no longer a significant improvement in accuracy performance. This phenomenon indicates that if a distributed buoy system is used for seabed anchor node calibration, having too many reference nodes will only lead to a linear increase in cost, but it is not conducive to improving accuracy.

\subsection{Influence of azimuth angle}

\begin{figure}[htbp]
	\centering
	\subfloat[]{
		\includegraphics[width=0.7\linewidth]{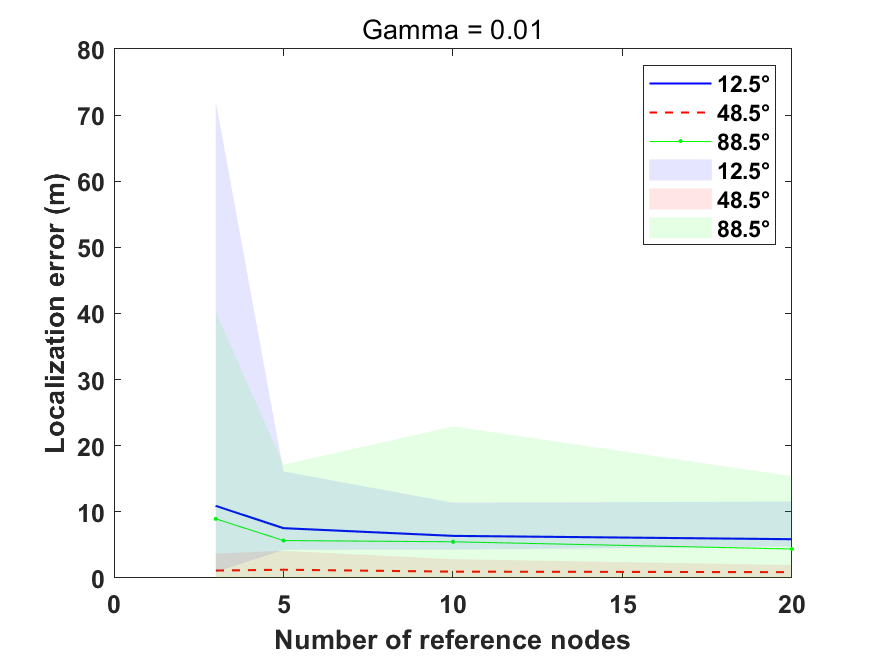}\label{fig08a}%
	}\\
	\subfloat[]{
		\includegraphics[width=0.7\linewidth]{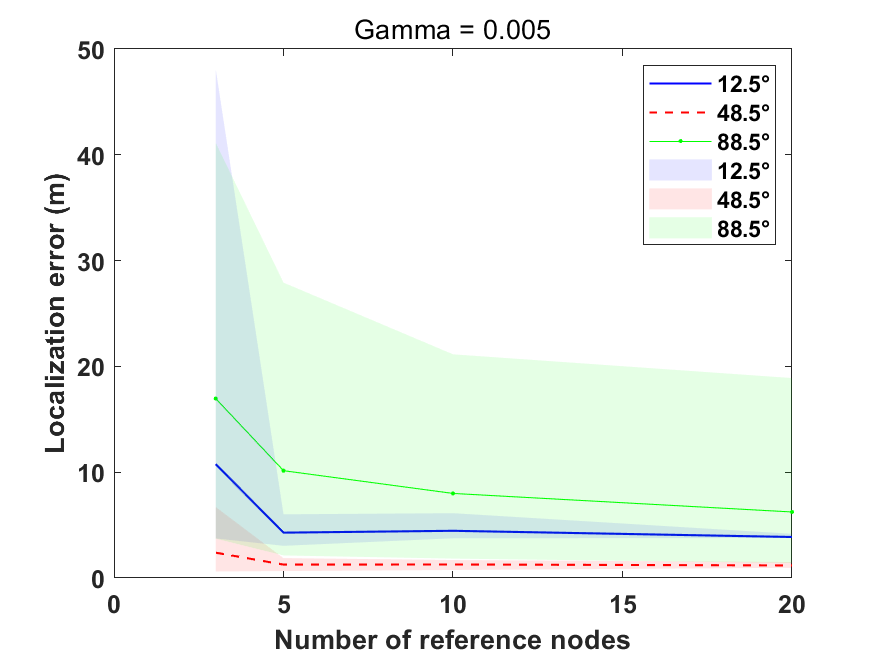}\label{fig08b}%
	}
	\caption{Localization error with randomly distributed reference nodes.}
	\label{fig08}
\end{figure}

\indent To test the influence of azimuth angle on localization accuracy performance, the localization error of anchor node with randomly azimuth angles of refernce nodes is given in Fig~\ref{fig08}, where Fig~\ref{fig08a} corresponds to $\gamma = 0.01$ and Fig~\ref{fig08b} corresponds to $\gamma = 0.005$. As the number of reference nodes increase, the reference nodes will have a higher probability of forming a uniform distribution, so the positioning error will decrease, especially when the number of reference nodes is less than 5.

\section{Real-World Experiments}

\begin{figure}[htbp]
	\centering
	\subfloat[Anchor node.]{
		\includegraphics[width=0.45\linewidth]{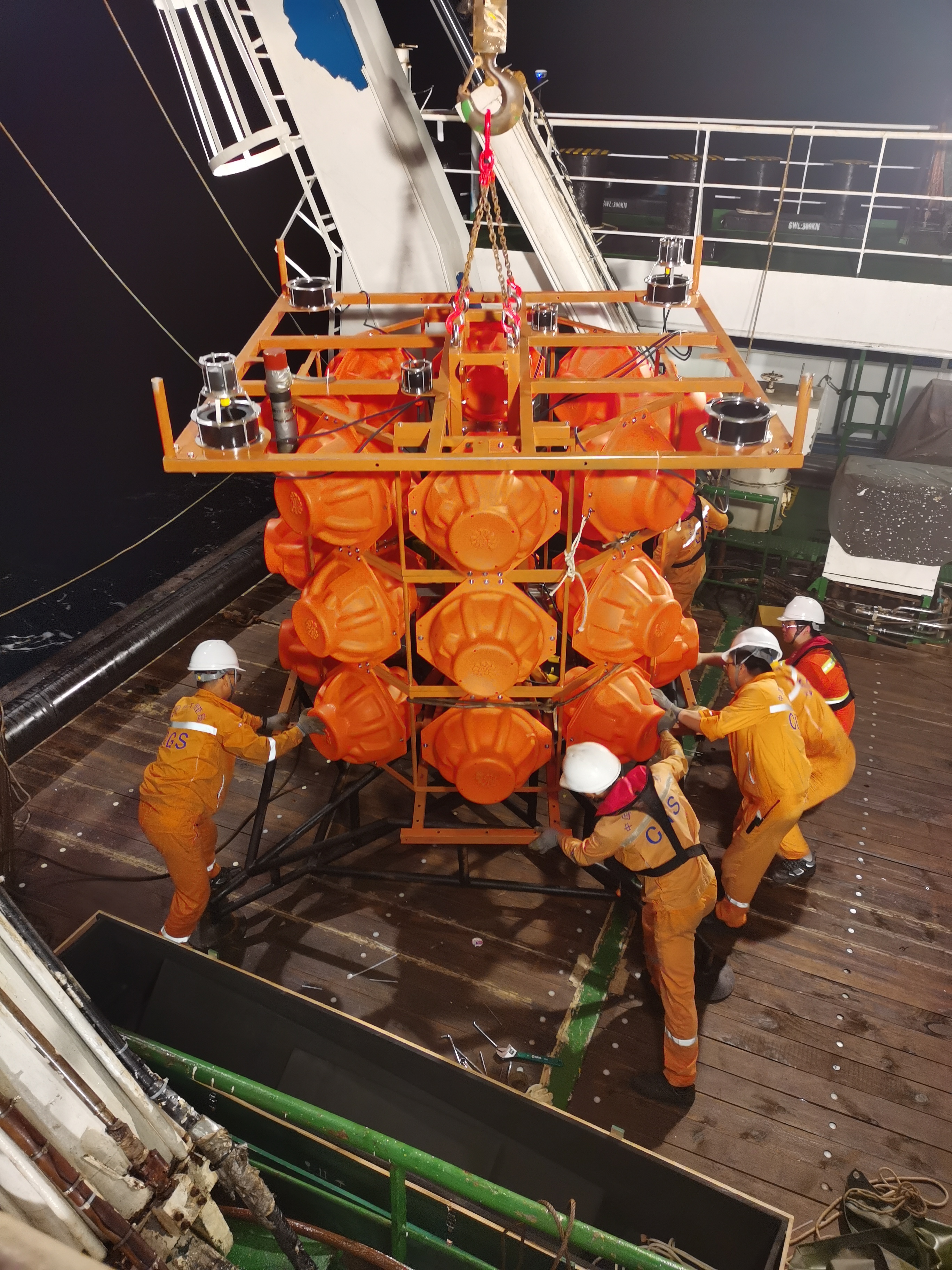}%
	}
	\subfloat[Ultra short base line unit.]{
		\includegraphics[width=0.45\linewidth]{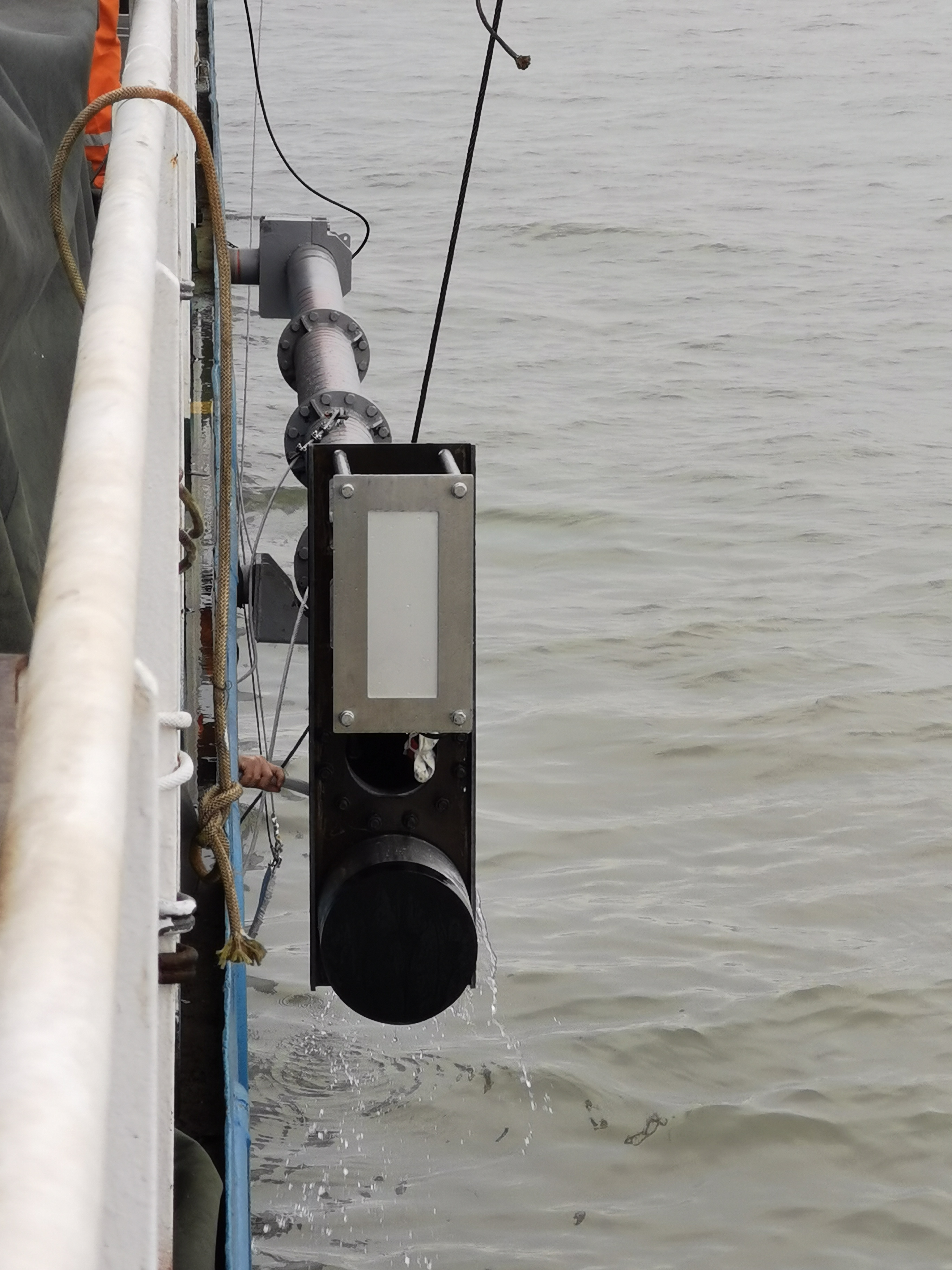}%
	}\hfill
	\subfloat[Relative azimuth of GPS and USBL in the horizontal direction.]{
		\includegraphics[width=0.8\linewidth]{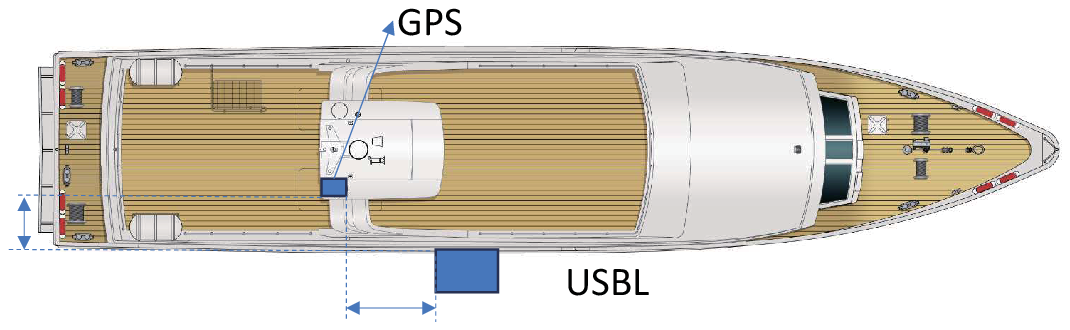}%
	}	
	\caption{Ship unit.}
	\label{fig09}
\end{figure}

To further evaluate the localization error of anchor node with different elevation angles, a deep ocean experiment was held in April 2023 in the South China Sea. There is a anchor node deployed on the seabed and a ship unit equipped with an ultra short baseline sonar system (USBL), which is shown in Fig.~\ref{fig09}. The real time location of the ship is aquired by global positioning system (GPS), then the real time position of USBL unit can be obtained through pole arm measurement. 

\begin{figure}[htbp]
	\centering
	\subfloat[Scenario of anchor node localization.]{
		\includegraphics[width=0.8\linewidth]{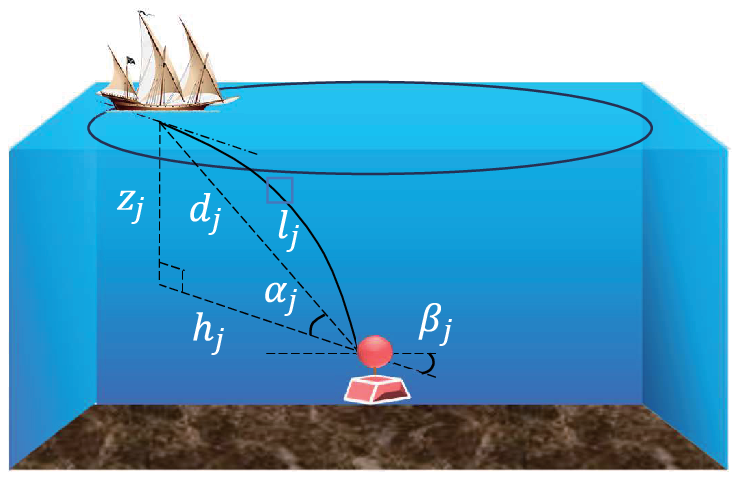}\label{fig10a}%
	}\hfill
	\subfloat[Topology of reference nodes and anchor node.]{
		\includegraphics[width=0.8\linewidth]{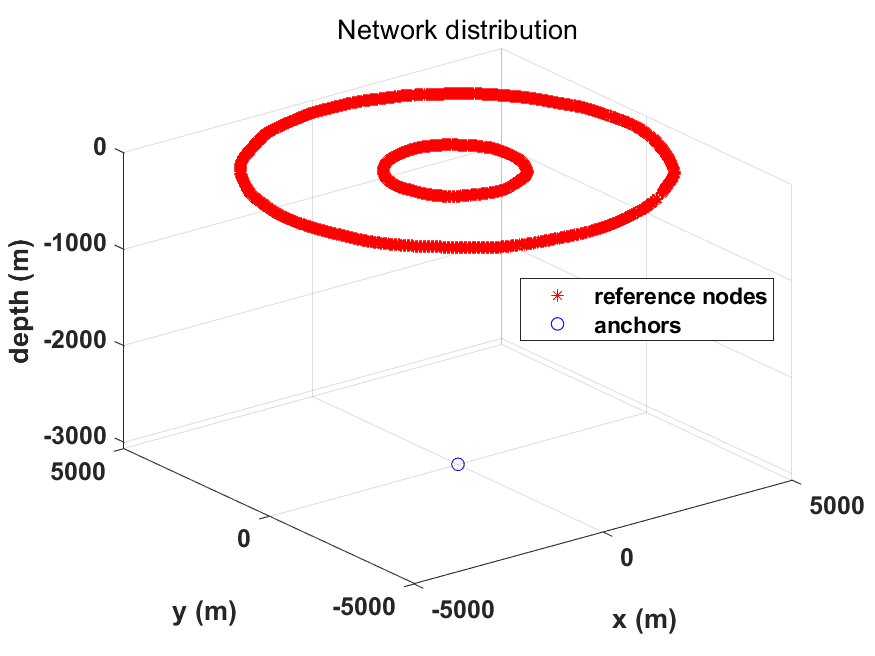}\label{fig10b}%
	}
	\caption{Experiment scenario.}
	\label{fig10}
\end{figure}

The scenario of anchor node localization is given in Fig.~\ref{fig10}. Duing the experiment, the ship unit sailed along a circular trajectory at a speed of 3 knots per second, while sending positioning request signals at fixed time intervals, then the anchor node reply answer signals when receiving positioning request signals. Through the round-trip signal interaction process, the TOA measurement data can be obtained. In this experiment, the ship navigates with two different circular radius to test the impact on positioning. The topology of reference nodes and anchor node is given in Fig.~\ref{fig10b}. The radius of the big circle is about $5000$ meters that is 1.5 times of the ocean depth, while the radius of the small circle is about $1600$ meters that is 0.5 times of ocean depth.

\begin{figure}[htbp]
	\centering
	\subfloat[CTD unit.]{
		\includegraphics[width=0.4\linewidth]{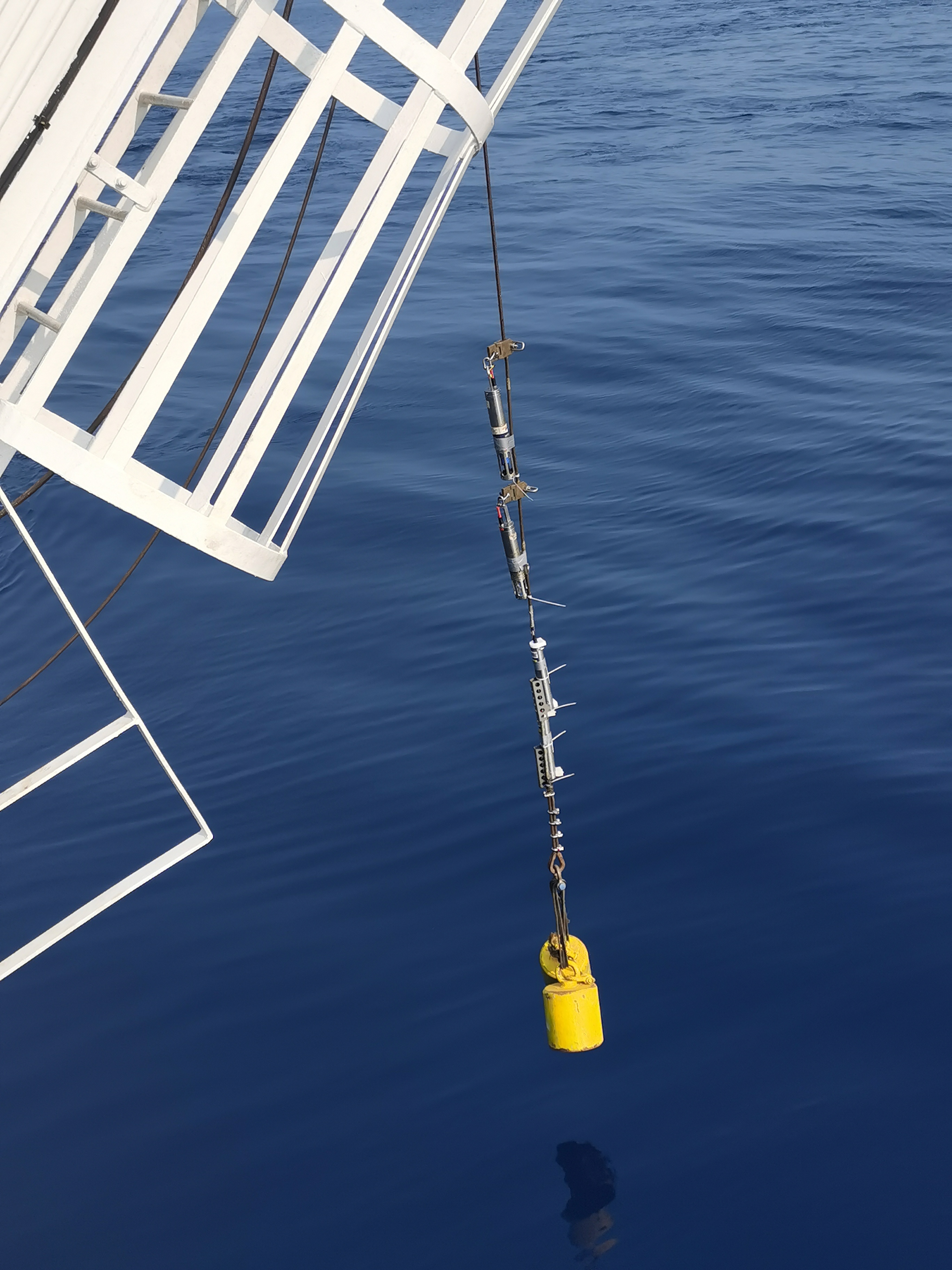}\label{fig11a}%
	}
	\subfloat[On-site SSP.]{
		\includegraphics[width=0.5\linewidth]{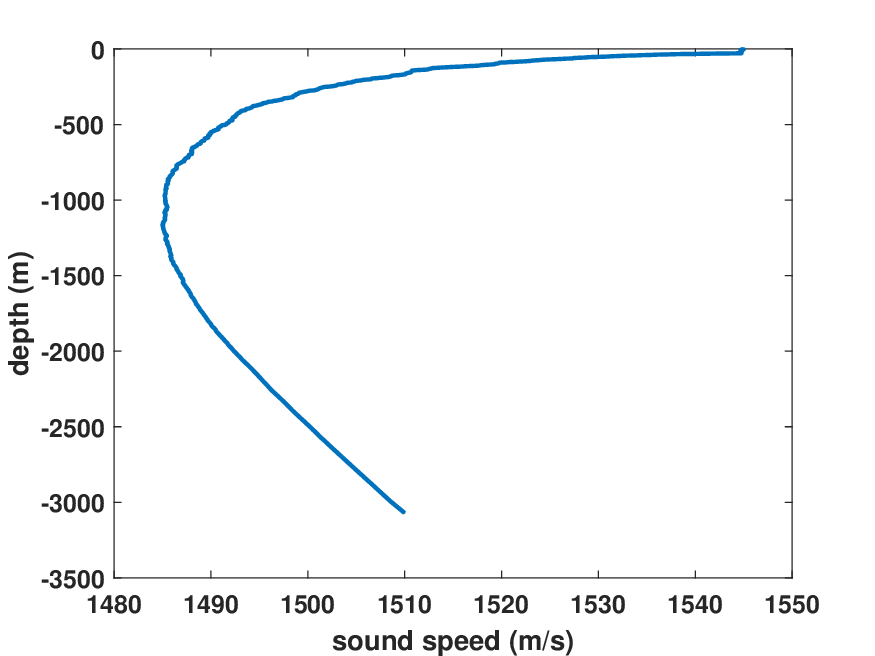}\label{fig11b}%
	}
	\caption{Measurement of sound speed.}
	\label{fig11}
\end{figure}

To obtain the on-site sound speed distribution, a conductivity, temperature, and depth (CTD) profiler is deployed into the ocean following the measurement of TOA data. The final measured sound speed distribution is shown in Fig.~\ref{fig11b}.

\begin{figure}[htbp]
	\centering
	\includegraphics[width=0.8\linewidth]{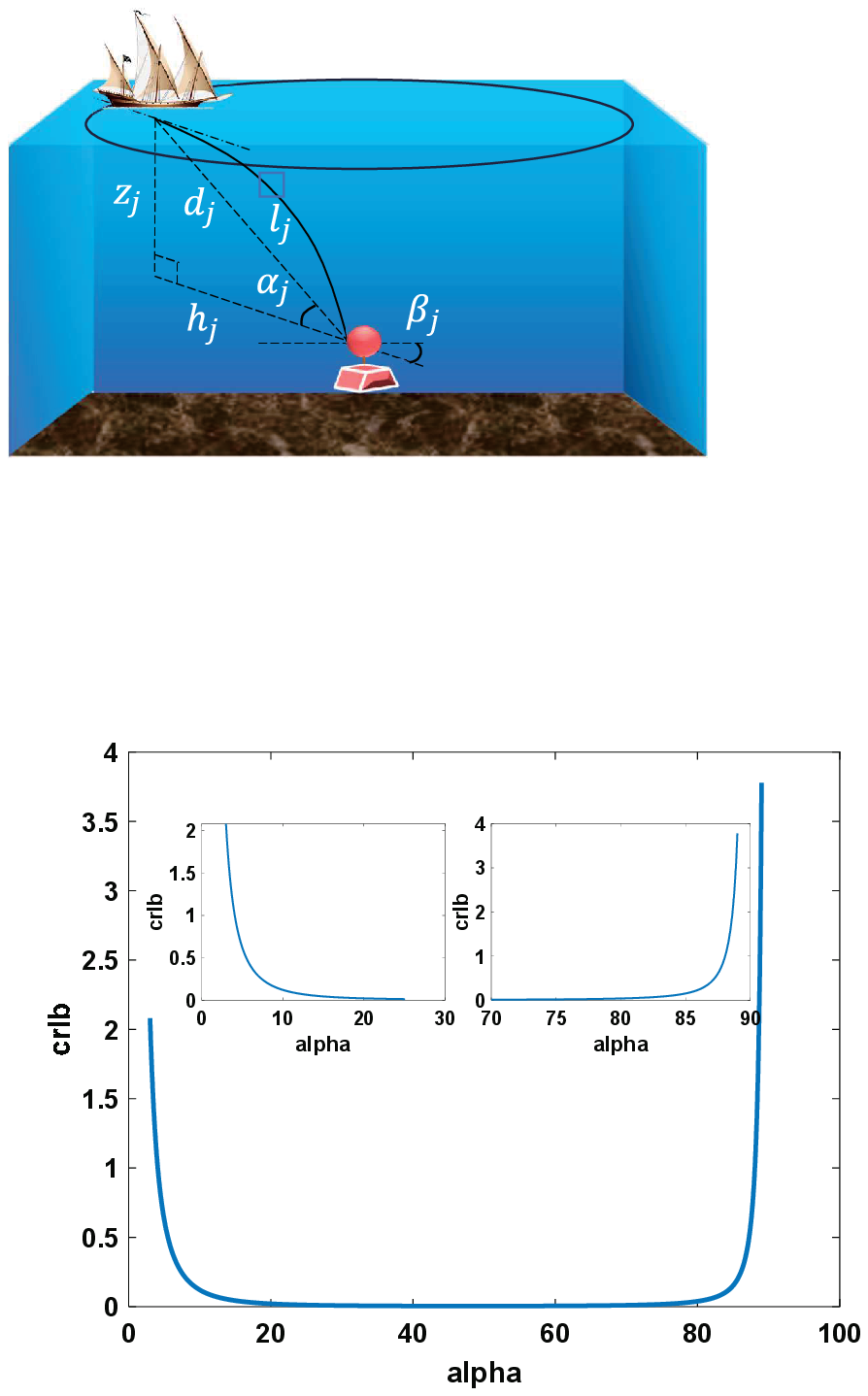}%
	\caption{$tr(CRLB)$ as a function of elevation angle $\alpha$ based on real sampled SSP.}
	\label{fig12}
\end{figure}

\indent The ocean depth is 3070 meters, and the radii of big and small circle are 5000 meters and 1600 meters, so the elevation angles can be computed as shown in Table~\ref{table3}. Then, the theoretical $tr(CRLB)$ can be computed by equation \eqref{eq29}. From theoretical analysis, the localization error of the small circle trajectory should be better than that of the big circle trajectory. The trend of $tr(CRLB)$ under real sample SSP is given in Fig.~\ref{fig12}.

\begin{table}[!htbp]
	\caption{$tr(CRLB)$ of two circle trajectory.\label{table3}}
	\centering
	\begin{tabular}{|c||c||c||c|}
		\hline
		trajectory & radius (m) & elevation angle ($^\circ$) & tr(CRLB)\\
		\hline
		small circle & 1600 & 62.4727 & 0.0072\\
		\hline
		big circle & 5000 & 31.5499 & 0.0083\\
		\hline
	\end{tabular}
\end{table}
	
\begin{figure}[htbp]
	\centering
	\subfloat[Localization error]{
		\includegraphics[width=0.7\linewidth]{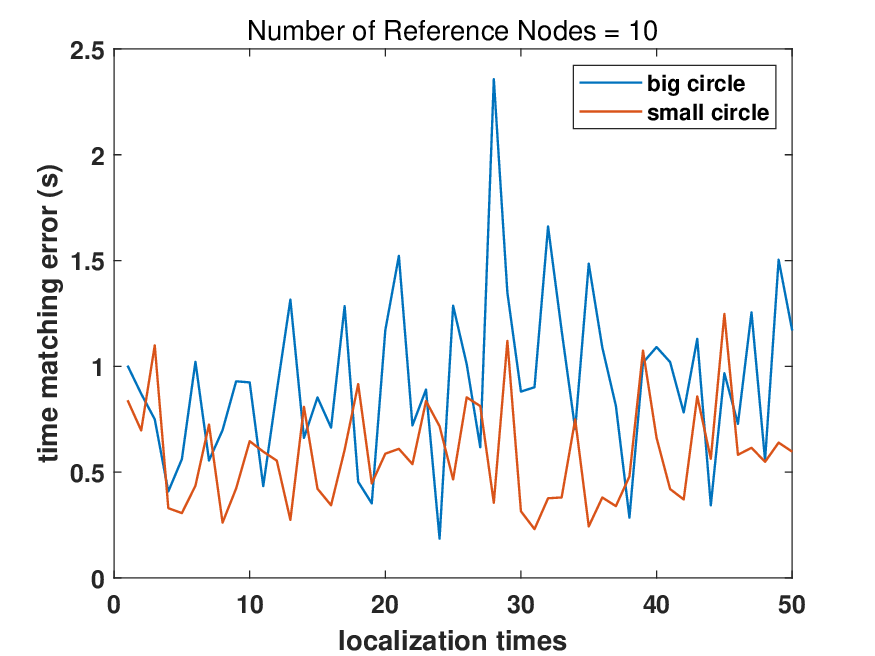}%
	}\hfill
	\subfloat[Average localization error.]{
		\includegraphics[width=0.7\linewidth]{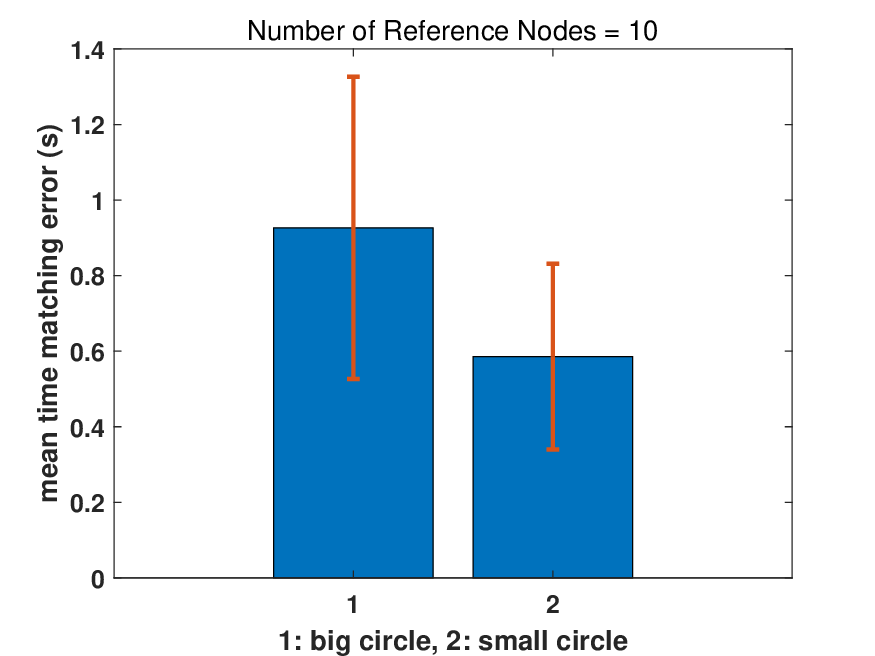}%
	}
	\caption{Ship unit.}
	\label{fig13}
\end{figure}

\indent To verify the analysis mentioned above, we conducted 50 times of localization on each trajectory with 10 randomly selected reference points. Since the real position of the seabed anchor node is not known, we set a time matching cost function as the basis for evaluating positioning accuracy as

\begin{equation}
	t_c = \sum_{j=1}^J (t_s - t_m)^2,
\end{equation}
where $t_c$ is the time cost, $t_s$ is the simulated signal propagation time, $t_m$ is the measured signal propagation time. The optimal positioning result should correspond to the minimum time cost. Thus, in this experiment, the small circle trajectory for positioning the anchor node should have better performance than the big circle. The positioning results are given in Fig.~\ref{fig13}, which is consistent with what we analysis before. To some extent, it confirms the theoretical findings in this paper.

\section{Conclusion}
In this paper, the optimal deployment of reference nodes for positioning seafloor anchor node is studied. The seafloor anchor node positioning scene is analyzed in detail and the deployment of reference nodes is optimized by optimizing the Fisher information matrix without constraints on the number of reference nodes and distance from the target to reference nodes. Meanwhile, the CRLB is derived. For the case with same measurement noise distribution among reference nodes, a semi-closed form optimal solution is derived, while for the case with different measurement noise distribution among reference nodes, it is proved that the optimal solution is equivalent to the situation where the error of each node is the same. Finally, simulation and ocean experimental results are consistent with the theoretical analysis, demonstrating the correctness of theoretical analysis.

\section*{Acknowledgments}
This document is the results of the research project funded by Natural Science Foundation of Shandong Province (ZR2023QF128), Laoshan Laboratory (LSKJ202205104), Qingdao Postdoctoral Science Foundation (QDBSH20220202061), National Natural Science Foundation of China (62271459), Fundamental Research Funds for the Central Universities, Ocean University of China (202313036).

\section*{Appendix A}
\indent Referring to Fig.~\ref{fig01}, the measured horizontal signal propagation distance will be
\begin{equation}
	\Delta \hat{h}_{i,j} = \frac{cos\theta_{i-1,j}}{\sin\theta_{i-1,j}}+n_{l,i,j}cos\theta_{i-1,j},\label{eq43}
\end{equation}
where $n_{h,i,j} = n_{l,i,j}cos\theta_{i-1,j} \sim N\left(0,\frac{\gamma^2 s_{i-1}^2\cos^2\theta_{0,j}}{s_0^2 - s_{i-1}^2\cos^2\theta_{0,j}}\right)$ is the measurement error in the horizontal direction. 

The total measured horizontal distance will be
\begin{equation}
	\hat{h}_{j} = \sum_{i=1}^I \Delta \hat{h}_{i,j} = h_{j} + n_{h,j},\label{eq44}
\end{equation}
where $n_{h,j} \sim N\left(0,\sum_{i=1}^I \frac{\gamma^2 s_{i-1}^2\cos^2\theta_{0,j}}{s_0^2 - s_{i-1}^2\cos^2\theta_{0,j}}\right)$

According to Fig.~\ref{fig01}, $\alpha_j$ and $\beta_j$ are the elevation and azimuth angles between the true target anchor node and the reference nodes, so we have
\begin{equation}
	\hat{d}_{j} = \frac{\hat{h}_{j}}{cos\alpha_{j}} = \frac{h_{j}}{cos\alpha_{j}} + \frac{n_{h,j}}{cos\alpha_{j}},\label{eq45}
\end{equation}
where $n_{d,j} = \frac{n_{h,j}}{cos\alpha_{j}}  \sim N\left(0,\sum_{i=1}^I \frac{\gamma^2 s_{i-1}^2\cos^2\theta_{0,j}}{(s_0^2 - s_{i-1}^2\cos^2\theta_{0,j})cos^2\alpha_{j}}\right)$. Therefore, we have 
$\sigma^2_{d,j} = \frac{cos^2\theta_{0,j}}{cos^2\alpha_{j}}\sum_{i=1}^I \frac{\gamma^2 s_{i-1}^2}{s_0^2 - s_{i-1}^2\cos^2\theta_{0,j}}$.

\section*{Appendix B}
For the jth reference node, the corresponding entries of \eqref{eq10} are calculated by

\begin{equation}
	\begin{split}
		\left.\frac{\partial \hat{d}_{j}}{\partial p_x}\right|_{\boldsymbol{p^t}} &= \frac{\partial \sqrt{(p^{t}_{x} - p^{r}_{x,j})^2 + (p^{t}_{y} - p^{r}_{y,j})^2 + (p^{t}_{z} - p^{r}_{z,j})^2 }}{\partial p^{t}_{x}}\\
		&= \frac{p^{t}_{x}-p^{r}_{x,j}}{d_j} = \cos\alpha_j \cos\beta_j.
	\end{split}
\end{equation}

\begin{equation}
	\begin{split}
		\left.\frac{\partial \hat{d}_{j}}{\partial p_y}\right|_{\boldsymbol{p^t}} &= \frac{\partial \sqrt{(p^{t}_{x} - p^{r}_{x,j})^2 + (p^{t}_{y} - p^{r}_{y,j})^2 + (p^{t}_{z} - p^{r}_{z,j})^2 }}{\partial p^{t}_{y}}\\
		&= \frac{p^{t}_{y}-p^{r}_{y,j}}{d_j} = \cos\alpha_j \sin\beta_j.
	\end{split}
\end{equation}

\begin{equation}
	\begin{split}
		\left.\frac{\partial \hat{d}_{j}}{\partial p_z}\right|_{\boldsymbol{p^t}} &= \frac{\partial \sqrt{(p^{t}_{x} - p^{r}_{x,j})^2 + (p^{t}_{y} - p^{r}_{y,j})^2 + (p^{t}_{z} - p^{r}_{z,j})^2 }}{\partial p^{t}_{z}}\\
		&= \frac{p^{t}_{z}-p^{r}_{z,j}}{d_j} = \sin\alpha_j.
	\end{split}
\end{equation}

\section*{Appendix C}
To prove \eqref{eq20}, we need to prove 
\begin{equation}
	\frac{(1-\cos^2\phi_1)}{\lambda} \geq 1,\frac{(1-\cos^2\phi_2)}{\lambda} \geq 1,\frac{(1-\cos^2\phi_3)}{\lambda} \geq 1
\end{equation}

Suppose by contradiction that $\frac{(1-\cos^2\phi_3)}{\lambda} < 1$. According to \eqref{eq18} and the properties of semi positive definite matrices $\boldsymbol{\Phi}$, there is $\left|\boldsymbol{\Phi}\right|>0$, which means $\lambda > 0$. Thus, the above inequality is equivalent to
\begin{equation}
	0 < 2\cos\phi_1\cos\phi_2\cos\phi_3 -\cos^2\phi_1-\cos^2\phi_2\label{eq50}
\end{equation}
However, because $\cos^2\phi_1+\cos^2\phi_2\geq2\cos\phi_1\cos\phi_2$, and $-1<\cos\phi_3<1$, it follows that
\begin{equation}
	\cos^2\phi_1+\cos^2\phi_2\geq2\cos\phi_1\cos\phi_2\cos\phi_3
\end{equation}
which contradicts \eqref{eq50}. Therefore $\frac{(1-\cos^2\phi_3)}{\lambda} \geq 1$, and similarly we have $\frac{(1-\cos^2\phi_2)}{\lambda} \geq 1,\frac{(1-\cos^2\phi_1)}{\lambda} \geq 1$.

Then, for the first term on the right side of \eqref{eq19}, we have
\begin{equation}
	\begin{split}
		\frac{(1-\cos^2\phi_3)}{|\boldsymbol{\hat{a}}|^2\lambda} \geq \frac{1}{|\boldsymbol{\hat{a}}|^2}
	\end{split}
\end{equation}
Similarly, we can obtain
\begin{equation}
	\frac{(1-\cos^2\phi_2)}{|\boldsymbol{\hat{b}}|^2\lambda} \geq \frac{1}{|\boldsymbol{\hat{b}}|^2}, 	
	\frac{(1-\cos^2\phi_1)}{|\boldsymbol{\hat{c}}|^2\lambda} \geq \frac{1}{|\boldsymbol{\hat{c}}|^2}
\end{equation}
This derivation can be similar to \cite{David2013OptimalPlacement,David2016OptimalPlacement,Xu2019OptimalTOA}. Considering the non-zero FIM determinant $(\lambda \neq 0)$, we have equation \eqref{eq20}.

\section*{Appendix D}
To simplify formula \eqref{eq25}, some special inequality relationships can be used. Assume $e,f \in \mathbb{R}$, and if $e>0$, $f>0$, there is 
\begin{equation}
	\frac{1}{e} + \frac{1}{f} \geq \frac{2}{\sqrt{e}\sqrt{f}}, \frac{1}{2\sqrt{e}\sqrt{f}} \geq \frac{1}{e+f}
\end{equation}
thus
\begin{equation}
	\frac{1}{e} + \frac{1}{f} \geq \frac{4}{e+f}
\end{equation}

The denominators of the first and second terms in \eqref{eq25} are considered as $e$ and $f$ so that

\begin{equation}
	\text{tr}(CRLB)	\geq \left(\sum_{j=1}^{J} \frac{\cos^2 \alpha_j}{4\sigma_d^2}\right)^{-1} + \left(\sum_{j=1}^{J} \frac{\sin^2 \alpha_j}{\sigma_d^2}\right)^{-1}
\end{equation}
and the equality is reached when
\begin{equation}
	\sum_{j=1}^{J} \frac{\cos^2 \alpha_j \cos^2 \beta_j}{\sigma_d^2}= \sum_{j=1}^{J} \frac{\cos^2 \alpha_j \sin^2 \beta_j}{\sigma_d^2}
\end{equation}
and \eqref{eq26} is satisfied.

\section*{Appendix E}
\indent Let
\begin{equation}
	\begin{split}
		P(\alpha)=\frac{4J\gamma^2\sum_{i=1}^{I}M_i(\alpha)-4J\gamma^2\cos^2\alpha\sum_{i=1}^{I}(M_i(\alpha))^2}{(J\cos^2\alpha)^2}\\
		-\frac{J\gamma^2\sum_{i=1}^{I}M_i(\alpha)+J\gamma^2\sin^2\alpha\sum_{i=1}^{I}(M_i(\alpha))^2}{(J\sin^2\alpha)^2}
	\end{split}\label{eq58}
\end{equation}
then $\frac{\partial P(\alpha)}{\partial \alpha}$ will be shown as equation \eqref{eq59}.

\setcounter{equation}{58}
\begin{figure*}
	\begin{equation}
		\begin{split}
			\frac{\partial P(\alpha)}{\partial \alpha} &= \frac{8J\gamma^2\sin2\alpha\sum_{i=1}^{I}\cos^2\alpha M_i(\alpha) - 8J\gamma^2\sin2\alpha\sum_{i=1}^{I}\cos^4\alpha M_i^2(\alpha)+ 8J\gamma^2\sin2\alpha\sum_{i=1}^{I}\cos^6\alpha M_i^3(\alpha)}{J^2\cos^8\alpha}\\
			&+ \frac{2J\gamma^2\sin2\alpha\sum_{i=1}^{I}\sin^2\alpha M_i(\alpha) + 2J\gamma^2\sin2\alpha\sum_{i=1}^{I}\sin^4\alpha M_i^2(\alpha)+ 2J\gamma^2\sin2\alpha\sum_{i=1}^{I}\sin^6\alpha M_i^3(\alpha)}{J^2\sin^8\alpha}
		\end{split}\label{eq59}
	\end{equation}
\end{figure*}

Since $\cos^2\theta_0/\cos^2\alpha \approx 1$, $M_i(\alpha) \approx \frac{s_{i-1}^2}{s_0^2-s_{i-1}^2\cos^2\theta_0}$. According to \eqref{eq4}, there is $s_{i-1}^2\cos^2\theta_0 = s_{0}^2\cos^2\theta_{i-1}$, where $0<\cos\theta_{i-1}<1$ because of $0<\theta_{i-1} < \frac{\pi}{2}$, so $s_0^2-s_{i-1}^2\cos^2\theta_0 > 0$, and $M_i(\alpha) > 0$. From another perspective, $M_i(\alpha) \approx \frac{\sigma^2_{d}}{\gamma^2}$, so it also holds $M_i(\alpha) > 0$.

\indent For \eqref{eq59}, the second part will always be positive, while for the first part, if $\cos^2\alpha M_i(\alpha) > 1$, then $\cos^6\alpha M_i^3(\alpha) - \cos^4\alpha M_i^2(\alpha) > 0$, if $0<\cos^2\alpha M_i(\alpha)< 1$, then $\cos^2\alpha M_i(\alpha) - \cos^4\alpha M_i^2(\alpha) > 0$. Therefore, there will always be $\sum_{i=1}^{I}\cos^2\alpha M_i(\alpha) - \sum_{i=1}^{I}\cos^4\alpha M_i^2(\alpha)+ \sum_{i=1}^{I}\cos^6\alpha M_i^3(\alpha) > 0$, meaning the first part of \eqref{eq59} is also positive. Finally, we obtain $\frac{\partial P(\alpha)}{\partial \alpha} > 0$. When $\alpha$ approaches $0$, $\cos\alpha \approx 1$ and $\sin\alpha$ goes to $0$, so $P(\alpha) < 0$ because the second part of \eqref{eq58} goes to $-\infty$, on the contrary, when $\alpha$ approaches $\frac{\pi}{2}$, $\sin\alpha \approx 1$ and $\cos\alpha$ goes to $0$, so $P(\alpha) > 0$ because the first part of \eqref{eq58} goes to $+\infty$. Based on the above analysis, the optimal solution exists and is unique.

\section*{Appendix F}
It shows that equation \eqref{eq38} has similar format with \eqref{eq30}, if $\alpha = a$ satisfies \eqref{eq30}, then $\alpha_j = a$ will also satisfies \eqref{eq38}. Similarly, we can obtain that $\alpha_{j+1} = a$ can also make $\frac{\partial \text{tr}(CRLB)}{\partial \alpha_{j+1}} = 0$.

\setcounter{equation}{60}
\begin{figure*}[!htbp]
	\begin{equation}
		\begin{split}
			\frac{\partial Q(\alpha_j)}{\partial \alpha_j} &= \frac{8\gamma^2\sin2\alpha_j\sum_{i=1}^{I}\cos^2\alpha_j M_i(\alpha_j) - 8\gamma^2\sin2\alpha_j\sum_{i=1}^{I}\cos^4\alpha_j M_i^2(\alpha_j)+ 8\gamma^2\sin2\alpha_j\sum_{i=1}^{I}\cos^6\alpha_j M_i^3(\alpha_j)}{\cos^8\alpha_j}\\
			&+ \frac{2\gamma^2\sin2\alpha_j\sum_{i=1}^{I}\sin^2\alpha_j M_i(\alpha_j) + 2\gamma^2\sin2\alpha_j\sum_{i=1}^{I}\sin^4\alpha_j M_i^2(\alpha_j)+ 2\gamma^2\sin2\alpha_j\sum_{i=1}^{I}\sin^6\alpha_j M_i^3(\alpha_j)}{\sin^8\alpha_j}
		\end{split}\label{eq61}
	\end{equation}
\end{figure*}

\indent Next, we will discuss if $\alpha_j = a$ is the only solution that satisfies \eqref{eq38}. Let
\setcounter{equation}{59}
\begin{equation}
	\begin{split}
		Q(\alpha_j) = \frac{4\gamma^2\sum_{i=1}^{I}M_i(\alpha_j)-4\gamma^2\cos^2\alpha_j\sum_{i=1}^{I}(M_i(\alpha_j))^2}{(\cos^2\alpha_j)^2}\\
		-\frac{\gamma^2\sum_{i=1}^{I}M_i(\alpha_j)+\gamma^2\sin^2\alpha_j\sum_{i=1}^{I}(M_i(\alpha_j))^2}{(\sin^2\alpha_j)^2}
	\end{split}
\end{equation}
then $\frac{\partial Q(\alpha_j)}{\partial \alpha_j}$ will be shown as equation \eqref{eq61}.

Similar to the analysis of \eqref{eq59}, we can obtain $\frac{\partial Q(\alpha_j)}{\partial \alpha_j} > 0$, and $Q(\alpha_j)< 0$ when $\alpha_j$ approaches $0$, while $Q(\alpha_j)> 0$ when $\alpha_j$ approaches $\frac{\pi}{2}$. So the optimal solution of $\alpha_j$ exists and is unique. Finally, we have $\alpha_1 = \alpha_2 = ... = \alpha_J = a$.

\bibliographystyle{IEEEtran}
\bibliography{IEEEabrv,draft_hw}

\section{Biography Section}

\vfill

\end{document}